\documentclass[12pt,preprint]{aastex}

\newcommand\teff{{T_{\rm eff}}}

\newcommand\delcol{$\Delta$(color)$_{\rm TO,RGB}$}

\newcommand\lta{\mathrel{\hbox{\raise 0.6 ex \hbox{$<$}\kern
                   -1.8 ex\lower .5 ex\hbox{$\sim$}}}}
\newcommand\gta{\mathrel{\hbox{\raise 0.6 ex \hbox{$>$}\kern
                   -1.7 ex\lower .5 ex\hbox{$\sim$}}}}
 
\shortauthors{Casagrande et al.}
\shorttitle{Testing $BV(RI)_C$ Transformations}
 
\begin{document}
 
\title{An Examination of Recent Transformations to the $BV(RI)_C$ Photometric
System From the Perspective of Stellar Models for Old Stars}

\author{Don A.~VandenBerg}
\affil{Department of Physics \& Astronomy, University of Victoria,
       P.O.~Bos 3055, Victoria, B.C., V8W~3P6, Canada}
\email{vandenbe@uvic.ca}

\author{L.~Casagrande}
\affil{Max Planck Institute for Astrophysics, Postfach 1317, 85741 Garching,
       Germany}
\email{Luca@MPA-Garching.mpg.de}

\author{Peter B.~Stetson}
\affil{Dominion Astrophysical Observatory, Herzberg Institute of Astrophysics,
       National Research Council, 5071 West Saanich Road, Victoria,
       BC~V9E~2E7, Canada}
\email{Peter.Stetson@nrc.gc.ca}

\begin{abstract}
Isochrones for ages $\gta 4$ Gyr and metallicities in the range $-2.5 \lta$
[Fe/H] $\lta +0.3$ that take the diffusion of helium and recent advances in
stellar physics into account are compared with observations in the
Johnson-Cousins $BV(RI)_C$ photometric system for several open and globular
star clusters.  The adopted color--$\teff$ relations include those which we
have derived from the latest MARCS model atmospheres and the empirical
transformations for dwarf and subgiant stars given by Casagrande et
al.~(2010, A\&A, 512, 54; hereafter CRMBA).  Those reported by VandenBerg \&
Clem (2003, AJ, 126, 778) have also been considered, mainly to resolve some
outstanding questions concerning them.  Indeed, for the latter, $V-I_C$ colors
should be corrected by $\approx -0.02$ mag, for all metal abundances, in order
to obtain consistent interpretations of the observed $(B-V,\,V)$-,
$(V-R_C,\,V)$-, and $(V-I_C,\,V)$-diagrams for M$\,$67 and the Hyades, as well
as for local subdwarfs.  Remarkably, when the subdwarfs in the CRMBA data set
that have $\sigma_\pi/\pi \le 0.15$ are superimposed on a set of 12 Gyr
isochrones spanning a wide range in [Fe/H], the inferred metallicities and
effective temperatures agree, in the mean, with those given by CRMBA to within
$\pm 0.05$ dex and $\pm 10$~K, respectively.  Thus the hot $\teff$ scale derived
by CRMBA is nearly identical with that predicted by stellar models; and
consequently, there is excellent consistency between theory and observations on
the H-R diagram {\it and} the different color-magnitude diagrams considered in
this investigation.  To obtain similar consistency, the colors obtained from
the MARC«S and VandenBerg \& Clem $(B-V)$--$\teff$ relations for metal-poor
dwarf stars should be adjusted to the red by 0.02--0.03 mag.  In general,
isochrones that employ the CRMBA transformations provide reasonably consistent
fits to our $BV(RI)_C$ photometry for main-sequence stars in the globular
clusters 47 Tuc, M$\,$3, M$\,$5, M$\,$92, and NGC$\,$1851 --- but not the 
cluster giants (when adopting the synthetic MARCS colors).  We speculate that
differences between the actual heavy-element mixtures and those assumed in the
theoretical models may be the primary cause of this difficulty.
\end{abstract}
 
\keywords{color-magnitude diagrams --- globular clusters: general ---
 globular clusters: individual (M$\,$3, M$\,$5, M$\,$92, NGC$\,$1851, 47 Tuc)
 --- Hertzsprung-Russell diagram --- open clusters and associations: general ---
 open clusters and associations: individual (Hyades, M$\,$67, NGC$\,$6791) ---
 stars: fundamental parameters (temperatures) --- stars: evolution}
 
\section{Introduction}
\label{sec:intro}
 
The empirical stellar $\teff$ scale is still uncertain by $\gta 100$~K in 
most parts of the H-R diagram despite painstaking spectroscopic and photometric
work by many investigators for many years (e.g., \citealt{gcc96};
\citealt{aam96}; \citealt{bsa02}; \citealt{rm05}; \citealt{naa07}).  Such
uncertainties have important consequences for the determination of other 
fundamental properties of stars --- notably their chemical abundances.  For
instance, the metallicities of solar neighborhood stars derived by Gratton et
al.~tend to be 0.1--0.25 dex more metal-rich than those reported by
\citet{cps07} because, in part, the temperatures adopted by the latter are up
to 150 K cooler than those estimated by the former (see \citealt{van08}, his 
Figs.~1, 2).  Even when the temperature is known to very high accuracy, as
in the case of the Sun, absolute abundances can vary by $\sim 0.2$ dex when 3D
hydrodynamical model atmospheres are employed instead of the classical 1D
hydrostatic models, departures from LTE are taken into account, improved atomic
and molecular data are incorporated into the analyses, etc.~(\citealt{ags05}).
Adding to the confusion is the fact that stellar models appear to have 
considerable difficulty matching the properties of globular cluster (GC) giants
as derived by \citet{cg97} when the same models reproduce quite well the $\teff$
and [m/H] values derived by the same researchers (\citealt{cgc00}) for the
Population II subdwarf standards (see, e.g., \citealt{bv01}, their Figs.~11--15;
\citealt{vbd06}, their Fig.~13).

[As shown later in this paper, significantly improved agreement between theory
and observations is obtained on the assumption of the recently revised
metallicity scale for GCs given by \citet{cbg09}.  It should be appreciated,
however, that their revision to lower [Fe/H] values (by typically $\sim 0.2$
dex) is due, in part (i.e., along with improvements to the spectra and
$\log\,{gf}$ values), to their adoption of lower temperatures to be consistent
with the \citet{aam99} $\teff$ scale for giants.  These temperatures may be
too low.  According to Casagrande et al.~(2010), the main difference between
their relatively high temperatures and those determined by \citet{aam96} for
dwarf and subgiant stars is the underlying absolute calibration of the Infrared
Flux Method.  Since a different calibration will mainly cause a zero-point
offset, we would expect that Alonso et al.~would obtain warmer $\teff$ values
for both dwarfs and giants were they to adopt the Casagrande et al.~calibration.
Thus, it is quite possible that Carretta et al.~{\it should} assume higher
temperatures, in which case their [Fe/H] estimates would also increase, thereby
moving them closer to the values originally published by \citet{cg97}.]

One obvious way of constraining the stellar $\teff$ scale is to obtain
photometry in many different bandpasses and then to examine the extent to which
a consistent interpretation of the data can be obtained on all of the possible
color-magnitude diagrams (CMDs) that can be constructed.  This approach 
motivated the studies by, in particular, \citet{vc03} and \citet{cvg04} of the
$BV(RI)_C$ and the Str\"omgren $uvby$ photometric systems, respectively.  Using
theoretical color indices derived from MARCS model atmospheres as the starting
point, these investigations determined the corrections that should be applied to
the synthetic colors in order to satisfy a variety of observational constraints.
Not surprisingly, the inferred corrections generally increased with decreasing
$\teff$ and they tended to be larger for colors involving ultraviolet or blue
magnitudes.  \citet{cbs07} and \citet{dcj08}, among others, have used the
resultant semi-empirical color transformations in their analyses of observed
CMDs with apparently quite favorable results. 

However, it is very difficult to avoid small zero-point or systematic errors in
{\it any} color--$\teff$ relations.  For instance, as discussed by VC03,
isochrones employing their transformations provide a good match to the Hyades
$[(B-V)_0,\,M_V]$- and $[(V-R),\,M_V]$-diagrams, on the assumption of
well-determined estimates of $(m-M)_0$, [Fe/H], and $Y$, but they tend to be
$\approx 0.02$ mag redder than the cluster observations on the
$[(V-I)_0,\,M_V]$-plane.  [Note that, $R$ and $I$ are used interchangeably 
with $R_C$ and $I_C$; i.e., all of the $R$ and $I$ photometry that is mentioned
in this paper is in the Cousins system, as defined by the standard stars of
\citet{gra82} and \citep[][1992]{lan83}.]  It was not at all clear to VC03
how best to explain this conundrum because no such difficulty was apparent when
they fitted isochrones to $BV$ and $VI$ data for the open clusters M$\,$67 and
NGC$\,$6791, or the very metal-deficient GC M$\,$68.  The subsequent study of
M$\,$67 by \citet{vs04} showed, in fact, that the \citet{mmj93} photometry used
in VC03's analysis agreed well with the CMDs produced by most, but not all,
other workers.  Only the \citet{san04} $VI$ observations were clearly different,
though (curiously) they provided the best match to the $(B-V)$--$(V-I)$ diagram
given by \citet{cca93}, based on their standardization of the Cousins system.
According to Sandquist, the main sequence of M$\,$67 is 0.01--0.03 mag bluer on
the $[(V-I)_0,\,M_V]$-plane than the determination by Montgomery et al.  If
correct (which we now believe to be the case, see \S~\ref{sec:obs}), this would
imply that the VC03 $(V-I)$--$\teff$ relations should be adjusted to the blue
by $\approx 0.02$ mag in order to achieve consistency with their $B-V$ and
$V-R$ transformations (and thereby also solve the Hyades problem).

In the meantime, one of us (PBS) has made considerable progress in his endeavor
(see \citealt{ste00}) to collect, reduce, and carefully calibrate to the
\citet{lan92} system a significant fraction of the world's photometry for open
and globular star clusters.  As shown later in this paper, his $BVI$ data for
M$\,$67 are in good agreement with those published by \citet{san04}, thereby
reinforcing our suspicion that the VC03 $(V-I)$--$\teff$ relations are too red
for near solar-abundance stars.  Moreover, he found that his current photometric
data (which are used here) for some GCs differ at the level of 0.01--0.03 mag
from published CMDs, sometimes in a systematic sense.  Besides the availability
of these very homogeneous data, there are two other recent developments that
make a further examination of color--$\teff$ relations worthwhile.

First, new and significantly improved grids of MARCS model atmospheres and
synthetic spectra have been published by \citet{gee08}.  In \S 2, we describe
how the latter have been processed into synthetic $BV(RI)_CJHK_S$ magnitudes.
(Note that a thorough study of the Str\"omgren color indices derived from the
new MARCS models is provided by \citealt{oge09}.)  Second, \citet{crm10} have
used the Infrared Flux Method (IRFM) to produce a new calibration of the $\teff$
scale for dwarf and subgiant stars spanning a wide range in [Fe/H] for which
the zero-point should be much more accurate than previous calibrations because
it is based on a number of solar twins.  Their results are presented in the
form of analytic expressions that relate many different photometric indices to
$\teff$ and [Fe/H].  As these color-temperature relations are based on field
stars, it is of some interest to examine how well they can reproduce the
main-sequence (MS) fiducials of star clusters when coupled with up-to-date
stellar evolutionary computations.  (The main advantage of such systems over
field stars is that their CMDs provide well-defined loci which connect stars
of the {\it same} metallicity.  In view of the mounting evidence for multiple
stellar populations in some GCs, one should focus on only those clusters with
very narrow fiducial sequences and, even in the most favorable cases, be wary
of the possibility that chemical abundance peculiarities may affect the fluxes
in some bandpasses more than others.) \S~3 presents an analysis of cluster and
field-star data using the MARCS, CRMBA, and VC03 transformations.  Finally,
a brief summary of the main conclusions of this investigation are given in \S 4.
[Because a companion paper by \citet{bsv10} focusses on the color--$\teff$
relations for the infrared, the present paper has been restricted, with one
exception, to a consideration of the $BV(RI)_C$ transformations.  In the case
of M$\,$67 we compare isochrones with $VK_S$ observations in order to
demonstrate the advantages of having  $V-K_S$ colors to complement those at
optical wavelengths.]

\section{Synthetic $BV(RI)_C$ Magnitudes Derived from MARCS (2008) Model
Atmospheres}
\label{sec:marcs}

In the following, synthetic colors and bolometric corrections\footnote{The
usual defintion of bolometric correction has been adopted: $BC_V=M_{\rm bol}
-M_V$, where $M_{\rm bol,\odot}= 4.75$.} in the Johnson-Cousins $BV(RI)_C$
system have been computed following the formalism described in \citet{cpf06}.
The only difference is the reference spectrum of Vega, now based on the updated
absolute spectrophotometry of \citet{boh07}, which intermingles {\it HST}
observations with model fluxes and provides the best accuracy available to
date, at least in the optical region.  Note that, despite the complications
posed by the pole-on and rapidly rotating nature of Vega, the effects on
the blue part of the spectrum are expected to be small or negligible
\citep[e.g.,][and references therein]{cpf06,boh07},
though some fine-tuning in the infrared may be necessary (see CRMBA). 
Briefly, the spectrum of Vega has been convolved with the $BV(RI)_C$ filter 
transmission curves of \citet{bes90b} and the results forced to match its 
observed magnitudes (\citealt{bes90a}) in order to determine the corresponding 
zero-points for each band.  The latter are needed to place onto the 
standard Johnson-Cousins system colors that are derived by convolving spectral 
libraries with the aforementioned filter transmission curves.

The new grid of MARCS synthetic spectra with ``standard" chemical abundances 
(i.e., $[\alpha/\rm{Fe}]=0.0$ for $[\rm Fe/H]\ge 0.0$, a linear increase of
$[\alpha/\rm{Fe}]$ from 0.0 at [Fe/H] $=0.0$ to 0.4 at [Fe/H $=-1.0$, and
$[\alpha/\rm{Fe}]=0.4$ for $[\rm Fe/H] \le -1.0$) has been used \citep{gee08}. 
This choice is well suited for the purpose of the present investigation, since 
the $\alpha$--enhancement measured in the majority of field stars and clusters 
follows this relation quite closely. It is known that most of the GCs seem to 
exhibit abundance variations and/or anomalies \citep[see e.g.][]{gsc04}, 
which, in principle, could also affect the predicted colors. This would require
a case-by-case study, tailoring synthetic models to the detailed chemical 
composition of each cluster, which is not always possible and clearly outside 
the scope of the present investigation.  (As shown later in this paper,
the inability of isochrones to reproduce the full CMD features of {\it some}
GCs may be telling us that the use of synthetic color--$\teff$ relations for
the standard mix of heavy elements might not be appropriate for these systems.)

For the sake of completeness, we generated synthetic colors using the (as yet, 
smaller) set of MARCS models having $[\alpha/\rm{Fe}]=0$ for all 
metallicities below solar.  Differences in the resultant $B-V$, $V-R$,
and $V-I$ colors amount to a few millimagnitudes (at most) for $\teff \gta
5000$~K, though they steadily increase with decreasing temperature, 
mostly because of the formation of molecules. Thus, a fine-tuning of the 
$\alpha$--enrichment could have a limited impact along the RGB or the lower 
MS, but the bulk of the CMD morphologies discussed in this paper is 
unaffected by this choice.  The full MARCS library is given for a
microturbulent velocity $\xi=2$ km/s.  Also, the geometry of the models
(plane-parallel for $\log\,g \ge 3.0$ or spherical for $\log\,g \le 3.5$)
have no significant impact on our broad-band colors in the overlap region,
as we have found from our tests that the two geometries produce a nearly
constant offset of a few millimagnitudes.  (Tables of synthetic magnitudes in
various photometric systems for all of these models will be published in a
forthcoming paper.)

[Interestingly, we found from two solar-like spectra for $\xi=1$ and 2 km/s that
this choice has a non-negligible ($\lta 0.02$ mag) effect on the calculated
$B-V$ color, while indices at longer wavelengths appear to be considerably less
affected.  This presumably occurs because microturbulence will partly
redistribute the flux in spectral regions that are crowded with lines (i.e.,
mainly in the blue).  This clearly introduces an additional degree of freedom
which can be avoided only by hydrodynamical simulations that treat the velocity
field in a consistent manner (\citealt{cat07}), though the impact of 3D model
atmospheres on synthetic colors is still largely unexplored (see
\citealt{cas09}; \citealt{klc09}).  Notwithstanding the fact that a
microturbulent velocity $\xi=1$ km/s is usually adopted for the Sun, the
generally good agreement between the synthetic (MARCS) and empirical (CRMBA)
color--$\teff$ relations reported in this paper suggests that $\xi=2$ km/s is
probably a good choice, at least for MS stars.]

Differently from other stars, the Sun does not provide a robust benchmark for 
testing synthetic colors.  In fact, it cannot be directly observed with the same
instrumentation used for stellar photometry; consequently, its colors can be
derived only indirectly.  Recent advances using photometry of solar twins
and solar spectrophotometry have improved upon this situation and suggest
that the latest MARCS model provides a decent fit to observations, especially
in terms of broad-band colors (always within $0.02$~mag), while larger
discrepancies at the bluest wavelengths and in intermediate-band filters
appear to be present \citep[][CRMBA]{edv08,mel10}.  The comparison presented 
in CRMBA also suggests that our adopted zero-points are indeed appropriate
for generating synthetic colors: the tendency of the MARCS solar spectrum to
return bluer than the ``observed" $B-V$ color index does, in fact, stem from
the model atmosphere for the Sun. 

There are a few other subtleties that are usually neglected when generating 
synthetic colors and we would like to comment on them since they could be 
relevant in the context of the present investigation (see also \citealt{bcp98} 
and \citealt{bes05} for a more detailed discussion).  In principle, to mimic
photometric measurements, synthetic photometry should reproduce the
instrumental system used and the same transformation equations determined
from observations should then be applied to replicate the standard system
under investigation.  In practice, this can hardly be achieved, especially if 
measurements from different instruments are used; whilst the zero-points of 
observational photometry are defined over an ensemble of well-measured stars,
the common practice in the case of synthetic photometry (as in this study) is
to set the zero-points using one reference star (usually Vega), for which its 
spectrophotometry and observed apparent magnitudes or color indices are well 
established.

Fortunately, there is a general good agreement between the highly 
standardized and homogeneous photometry used in this paper for open and 
globular clusters and other independent measurements, despite the known small 
differences between the \citet{lan92} and Cousins standards \citep{bes95}. This
means that transformations from the instrumental to the standard system are
indeed accurate and reproducing the latter using only Vega returns meaningful 
synthetic colors. Nevertheless, a contemporary standard system, although well 
defined by a list of standard stars, might not represent a real linear system 
anymore, implying that is impossible to realize it with a unique passband and
a linear transformation. Therefore, other than the passband, when trying
to reproduce a given set of observations, the same linear and non-linear
transformations used to place observations onto the standard system should
be adopted.  In practice, this task is very difficult to achieve; as a result,
corrections of a few percent to the synthetic colors cannot be totally excluded
across the whole temperature range of the models.  Despite this discouraging
scenario, the general good agreement between observed and synthetic colors
shown in the next section of this paper suggests that both are well
standardized, and hence that the comparison between observed stars and 
theoretical isochrones can be used to gain insights concerning synthetic 
color--$\teff$ relations.

Summarizing: from the above discussion, it should be kept in mind that model 
inaccuracies as well as detailed chemical composition and photometric
uncertainties may all play a role in the final results and it is generally
quite difficult to disentangle them; consistency in different bands at the
level of 0.01-0.02 mag can therefore be regarded as excellent.  In the
following, these possibilities have been taken into account and they are
discussed when it is relevant to do so.

\section{Star Cluster and Field Subdwarf Constraints on the $BV(RI)_C$
Transformations}
\label{sec:obs}

In the following analysis, theoretical isochrones will be transformed to the
observed plane using the CRMBA, MARCS, and VC03 color--$\teff$ relations and
then compared with the CMDs for a few open and globular star clusters and field
subdwarfs that span a wide range in [Fe/H].  Most of the models are taken from
the new Victoria-Regina grids that have been generated by D.~VandenBerg et
al.~(2010, in preparation; hereafter VR2010), though some of them have been
computed specifically for this or other projects currently underway.
Nevertheless, the same version of the Victoria evolutionary code has been used
in all instances.

Because this code has undergone substantial revisions since the last
presentation of Victoria-Regina isochrones (VandenBerg et al.~2006), we
provide a brief summary of the main modifications that have since been made
to it.  First, the latest rates for the $pp$-chain and the CNO-cycle (see,
e.g., \citealt{wei08}), including, in particular, that for the important
$^{14}$N(p,$\gamma$)$^{15}$O reaction (\citealt{mfg08}) have been adopted.
Second, the gravitational settling of helium (and lithium), as well as
turbulent mixing below envelope convection zones are now treated using methods
very similar to those described by \citet{pm91}.  (The observed solar Li
abundance was used to constrain the amount of mixing in a Standard Solar
Model, and thereby to determine the value of the free parameter in our adopted
formulation of this additional mixing; see the VR2010 paper for details.)
Third, we have implemented the improved conductive opacities reported by
\citet{cpp07}.  Fourth, the model atmospheres that are needed to define the
outer boundary conditions for the stellar interior models were obtained by
integrating the hydrostatic equation in conjunction with the scaled empirical
\citet{hm74} $T$--$\tau$ relation given by \citet{vp89}.  As shown by
\citet{vee08}, this choice provides a very good approximation to the use of
scaled-solar, differentially corrected MARCS model atmospheres as boundary
conditions over wide ranges in $\teff$, $\log\,g$, and metallicity.  (Indeed,
this paper contains quite a thorough discussion of not only the impact of
different treatments of the atmospheric layers on the predicted temperature
scale of stellar models, but also of the associated undertainties.  
Interested readers are encouraged to refer to this work.)
 
In what follows, we will show that the VR2010 models satisfy all of the
observational constraints that have been considered rather well.  It is, of
course, quite possible that the good consistency we have bound between theory
and observations is fortuitous to some extent; i.e., errors in one or more
aspects of the models or observations have compensated for errors in other
factors that play a role to give a misleadingly rosy picture.  The models
still employ the crude mixing-length theory of convection and a very {\it ad
hoc} prescription for turbulent mixing, for instance; consequently, the 
physics incorporated in them can certainly be improved upon.  Nevertheless,
the Victoria-Regina models, coupled with the CRMBA or MARCS color--$\teff$
relations, appear to pass the tests to which they have been subjected (so far).
Accordingly, one can have considerable confidence in the results that are
obtained when these models are used, say, to interpret stellar populations
data.\footnote{The referee queried how well the isochrones computed by other
workers fare in similar comparisons, which is tantamount to asking how well
the VR2010 models agree with those published by other groups.  We have not
attempted to carry out such an analysis, which is beyond the scope of the
present work, and which is worth doing only if it is first demonstrated that
the evolutionary tracks produced by the different codes in use today are in
good agreement when very similar physics is assumed.  As shown by \citet{wei07},
who carried out such experiments, it is not straightforward to obtain the level
of agreement that one would like (and expect) for, in particular, the predicted
lifetime of a star of a given mass and chemical composition.
In any case, we can report that both the computed track and the predicted age at
the RGB tip from the Victoria code for the 
$1.0 {{\cal M}_\odot}$, $Y=0.28$, $Z=0.02$ test case considered by Weiss et
al.~are within 1--2\% of those obtained from the BASTI code (\citealt{pcs04}).
Similar good agreement has been found when comparisons are made with the tracks
produced by the MESA code (B.~Paxton et al., in preparation; also see
http://mesa.sourceforge.net) using more up-to-date physics.  These and other
tests of the reliability of the Victoria-Regina models are reported in much
greater detail in the VR2010 study.}

Turning to the observations: where possible, distances derived from
{\it Hipparcos} parallax measurements are assumed, along with current best
estimates of the basic stellar/cluster parameters.  However, even when (for
instance) the adopted distance moduli are uncertain by $\sim 0.1$--0.2 mag, as
in the case of most of the star clusters considered here, such uncertainties
are of little importance.  Even if the isochrones do not fit the observations
particularly well in an absolute sense (for whatever reason), any discrepancies
that are found {\it should} be apparent in the many different CMDs that can be
constructed from $BV(RI)_C$ photometry {\it if} the color transformations that
are used lead to a similar and consistent interpretation of the data on all
color-magnitude planes.  This does require, of course, that the observations
are themselves free of systematic errors and that the extinction in each of
the filter passbands is accurately determined if there is significant foreground
reddening.  (As already noted, the possibility that chemical abundance anomalies
may affect some color indices more than others is also a concern.)

Our examination of color--$\teff$ relations for stars having [Fe/H] $\gta 0.0$
will focus on the Hyades, M$\,$67, and NGC$\,$6791 open clusters.  Those
appropriate to metal-deficient stars will be assessed using $\sim 100$ nearby
subdwarfs having [Fe/H] $\lta -0.5$ (of which nearly three dozen have
$\sigma_\pi/\pi\lta 0.15$), as well as the globular clusters 47 Tucanae,
NGC$\,$1851, M$\,$5, M$\,$3, and M$\,$92, which have metallicities in the range
$-0.8\gta$ [Fe/H] $\gta -2.4$.  Unless noted otherwise, the most recent
calibrations of the cluster photometry in the \citet{ste00} database are used
in this study.  Unfortunately, $R_C$ photometry is available only for the
Hyades, NGC$\,$1851, M$\,$5, M$\,$92, and most of the field subdwarfs.  For the
other objects, our analysis is necessarily restricted to $BVI_C$ photometry.

\subsection{The Hyades ([Fe/H] $\approx +0.14$)}
\label{subsec:hyades}

The Hyades provide an especially powerful constraint on stellar models because
all of its basic parameters are known to high precision.  It has $E(B-V) = 0.0$,
$(m-M)_0 = 3.334 \pm 0.024$ from {\it Hipparcos} (\citealt{vl09}), [Fe/H]
$= +0.14 \pm 0.03$ from high-resolution spectroscopy (e.g., \citealt{ccl97};
also see \citealt{psc03}), and $Y \approx 0.26\pm 0.005$ from the cluster
binaries (\citealt{lfj01}; VC03).  As discussed in the VR2010 study, a Standard
Solar Model --- one that reproduces the properties of the Sun
at its present age --- requires
$Y_{\odot,{\rm initial}} = 0.26575$ if $Z_\odot = 0.01652$, assuming the solar
heavy-element abundances given by \citet{gs98}, and a value of 2.01 for the
mixing-length parameter, $\alpha_{\rm MLT}$.\footnote{Although the VR2010
investigation also provides evolutionary tracks and isochrones
for the solar distribution of the
metals determined by \citet{ags05}, these models are not used in this
investigation.  This is mainly for the reason that recent revisions to nuclear
reaction rates [especially for $^{14}$N$(p,\,\gamma)^{15}$O --- see
\citealt{mfg08}] imply a significant increase ($\gta 0.06 {{\cal M}_\odot}$) in
the mass of the lowest mass star that has a convective core at central H
exhaustion (\citealt{cd09}), thereby reducing the age of the oldest isochrone
that predicts the existence of a gap near the turnoff.  While it may still be
possible to produce models for the Asplund et al.~metal abundances that possess
a gap at the observed luminosity by allowing for sufficient convective core
overshooting, fine-tuning of the treatment of overshooting seems to be necessary
to achieve this, especially if diffusive processes are also treated (see the
careful and thorough analysis of this problem by \citealt{msw10}).  In view of
the additional difficulties presented by the Asplund et al.~metals mixture
for helioseismology (e.g., \citealt{bbs05}), it seems advisable to use stellar
models that assume the relative abundances of the heavy elements tabulated by
\citet{gs98}.}  This value of $Z_\odot$, together with the derived values of
$Y$ and [Fe/H] for the Hyades, imply that the cluster stars have $Z \approx
0.023$.  In fact, the VR2010 models provide a superb fit to
the mass-$M_V$ relation for the binaries (not shown here because it is
essentially identical to that shown by VC03; see their Fig.~21) if they assume
$Y = 0.257$, $Z = 0.023$, and an age of $\approx 700$ Myr.

\begin{figure}
\plotone{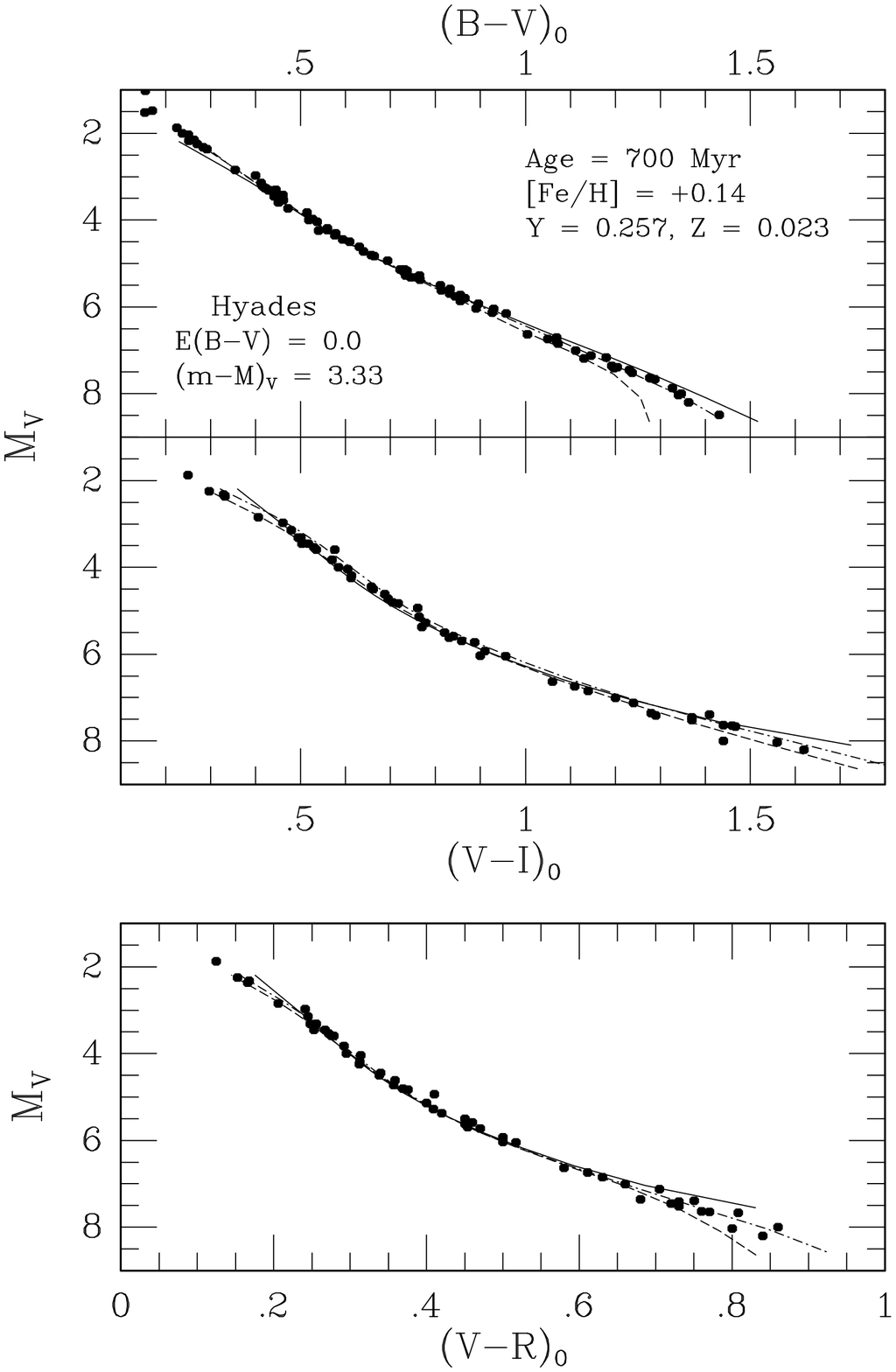}
\caption{Comparison of an isochrone for the indicated age and chemical
abundances with photometry of the Hyades on three different color-$M_V$ diagrams
assuming the CRMBA, MARCS, and VC03 color transformations (solid, dashed, and
dot-dashed curves, respectively) and $(m-M)_V = 3.33$.  The sources of the
photometry are mentioned in the text.}
\label{fig:fig1}
\end{figure}

Figure~\ref{fig:fig1} illustrates how well the main-sequence segment of an
isochrone for these parameter values reproduces the Hyades CMDs (see VC03)
constructed from $BV$, $VR$, and $VI$ photometry reported by \citet{tj85},
\citet{jt88}, \citet{rei93}, and \citet{dhd01}, when the three different sets of
color--$\teff$ relations considered in this investigation are assumed.  Stars
having $B-V \lta 0.9$, $V-R \lta 0.7$, and $V-I \lta 1.5$ (which corresponds,
in turn, to values of $\teff\gta 5100$, 4450, and 4200~K) are well-fitted by
the models and there is very good consistency between the three loci, except
on the $(V-I),$\,$M_V$ diagram, where the VC03-transformed isochrone is $\approx
0.02$ mag too red.  At faint magnitudes, the CRMBA colors tend to be slightly
too red, while the MARCS transformations yield $B-V$ and $V-R$ colors, but not
$V-I$ indices, that are too blue.  The fact that the observed $V-I$ colors are
so well reproduced using the MARCS transformations while the colors derived
from bluer bandpasses (notably $B$) tend to become discrepant at lower values
of $\teff$ suggests that the MARCS atmospheres for cool, super-metal-rich stars
have insufficient blanketing at shorter wavelengths.  If this suspicion is
correct, then it is the synthetic $B-V$ and (to a lesser extent) $V-R$ colors
that need to be corrected in order to achieve better consistency between the
three color planes at faint magnitudes --- more so than the isochrone
temperatures.  The only obvious problem with the VC03 transformations appears
to be the aforementioned offset in the $V-I$ colors.  Whether the discrepancies
between the solid curve and the observations is an artefact of the analytic
expressions used by CRMBA to represent their color--$\teff$
relations\footnote{Although the equations presented by CRMBA are in the form
that is traditionally used to describe such results, a different functional
relationship appears to be needed for M dwarfs (see \citealt{cfb08}), which
could alleviate the problem discussed here.  However, except for solar
abundance stars, it is not yet possible to extend the CRMBA calibrations to
very red colors primarily because of the limited metallicity range encompassed
by nearby M dwarfs.  Indeed, the analytic expressions given by CRMBA are valid
only for the color ranges that are specified in their paper.} or an indication
of, say, a small problem with the model temperatures is not clear.

\subsection{M$\,$67 ([Fe/H] $\approx 0.0$)}
\label{subsec:m67}

\begin{figure}
\plotone{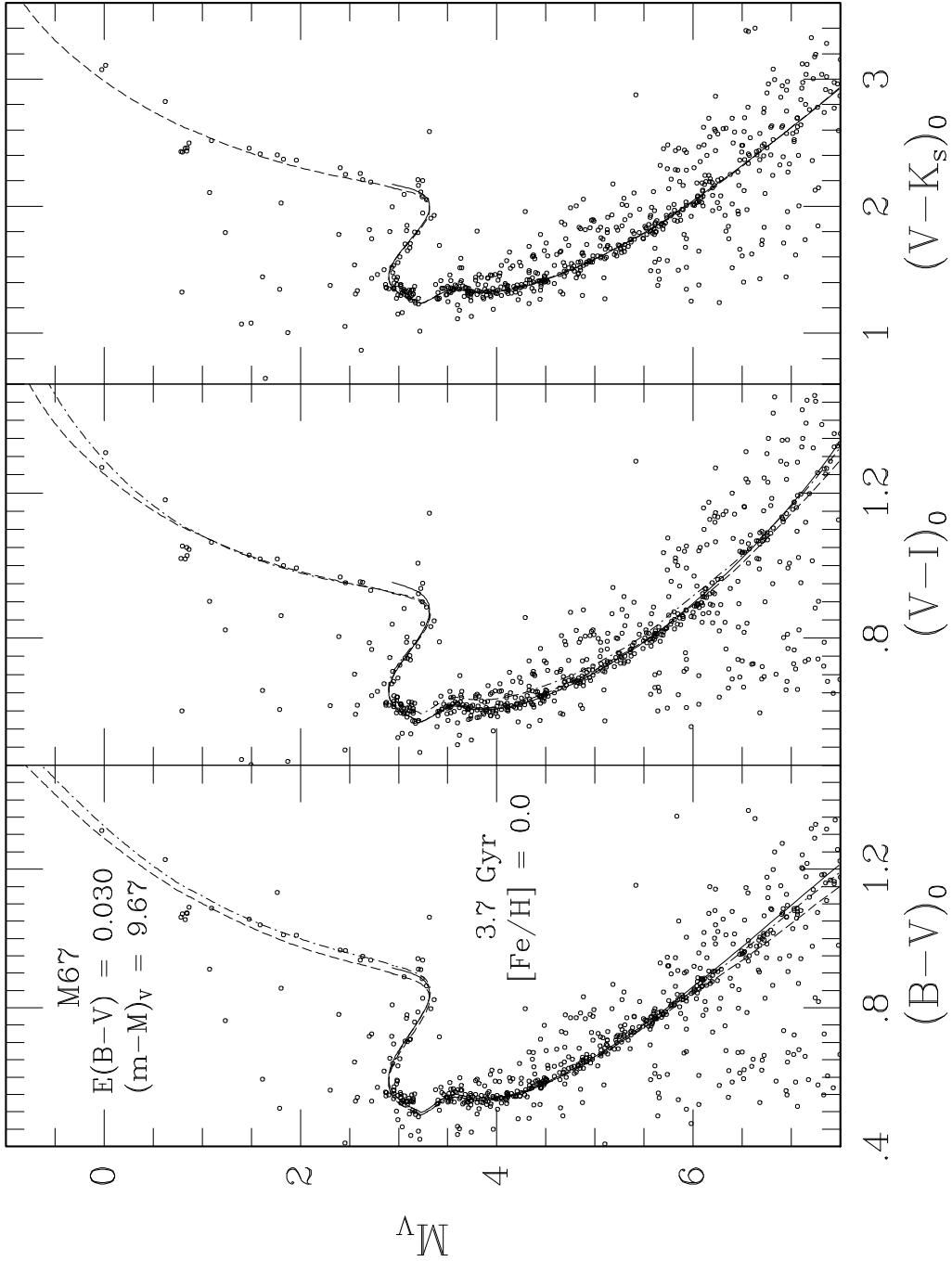}
\caption{Similar to the previous figure, except that a 3.7 Gyr isochrone for
the solar metallicity (from \citealt{mrr04}) has been overlaid onto three
different CMDs for M$\,$67.  The solid, dashed, and dot-dashed curves represent,
in turn, the assumption of the CRMBA, MARCS, and VC03 color transformations.
The latter predict $V-I$ colors that are $\sim 0.02$ mag too red.  Note that
the $BVI$ photometry plotted here was newly reduced by one of us (PBS). }
\label{fig:fig2}
\end{figure}

According to the results of high-resolution spectroscopy, M$\,$67 has [Fe/H]
$=0.0\pm 0.03$, with very close to solar m/H number abundance ratios for all
of the most important heavy elements (\citealt{tet00}; \citealt{rsp06}).
Current best estimates of the foreground reddening favor values in the range
$0.03\lta E(B-V) \lta 0.04$ (\citealt{ntc87}; \citealt{sfd98}; \citealt{svk99}).
In the course of examining the $BVI$ photometry reduced by one of us
(PBS), along with 2MASS near-infrared observations (\citealt{scs06}), we found
that it was possible to obtain very nice consistency of a 3.7 Gyr isochrone for
the solar metallicity with all of the available data if $E(B-V) = 0.03$ was
adopted.  Since $E(V-K_S) = 2.72\,E(B-V)$ (\citealt{mcc04}), even a change of
0.01 mag can have noticeable consequences for the consistency, or not, of
optical and near-IR CMDs.  If this reddening is assumed, together with an
apparent distance modulus $(m-M)_V = 9.67$, we obtain the comparison between
theory and observations shown in Figure~\ref{fig:fig2}.  As in the case of the
Hyades, the different line types are used to represent the different color
transformations that have been used.  Note that the true distance modulus
assumed here, $(m-M)_0 \approx 9.58$, is in excellent agreement with the values
of 9.57 and 9.60 derived by Sarajedini et al.~and \citet{san04}, respectively.

The isochrone which appears in this figure is the same one that was fitted to
the CMD of M$\,$67 by \citet{mrr04}.  It assumes the heavy-element abundances
given by \citet{gs98} and it takes the gravitational settling and radiative
accelerations of helium and the metals into account.  As also demonstrated by
Michaud et al., this isochrone reproduces the morphology of the cluster CMD
in the vicinity of the turnoff, including the location of the gap, particularly
well.  (Michaud et al.~did not need to assume any convective core overshooting
in their models, though some overshooting would almost certainly be required to
compensate for the recent revisions to nuclear reactions; see footnote 3.)
Because that isochrone terminated just above the base of the red-giant
branch (RGB), the evolution to higher luminosities has been represented by the
giant branch segment from a non-diffusive evolutionary track for 1.40 solar
masses and the same chemical abundances.  This locus, which was computed using
the VR2010 code, had to be shifted by only $\delta\log\teff = 0.0022$ in order
to achieve continuity with the Michaud et al.~isochrone on the theoretical
plane.
     
Fig.~\ref{fig:fig2} shows that the CRMBA, MARCS, and VC03
$B-V$ and $V-I$ transformations are in very good agreement from $\sim 2$ mag
below the turnoff up to the base of the RGB, except that the VC03 $V-I$ colors
apparently suffer from a small, nearly constant, offset.  Only on the
$[(B-V)_0,\,M_V]$-diagram are there some noticeable differences in a systematic
sense.  The CRMBA $B-V$ colors for cool stars ($\teff\lta 4950$ K, $(B-V)_0\gta
0.92$) appear to be slightly too red (possibly a reflection of the limitations
of the analytic expression used to describe these color transformations, as
already mentioned in footnote 4), while those based on MARCS model atmospheres
cause the isochrone to deviate to the blue side of both the observed RGB and
the lower main sequence, reminiscent of our findings in the case of the Hyades.
The best fit to the cluster $BV$ photometry is obtained using the VC03
transformations (by design, since the colors predicted by previous generations
of MARCS model atmospheres were corrected so that isochrones would reproduce
the {\it slopes} of cluster main sequences on the different color planes).  For
the latter to provide a fully consistent explanation of the $VI$ data, the VC03
$(V-I)$--$\teff$ relations for dwarf stars should be adjusted to the blue by
the amounts needed to bridge the gap between the dot-dashed curve and the
observed MS of M$\,$67.

\begin{figure}
\plotone{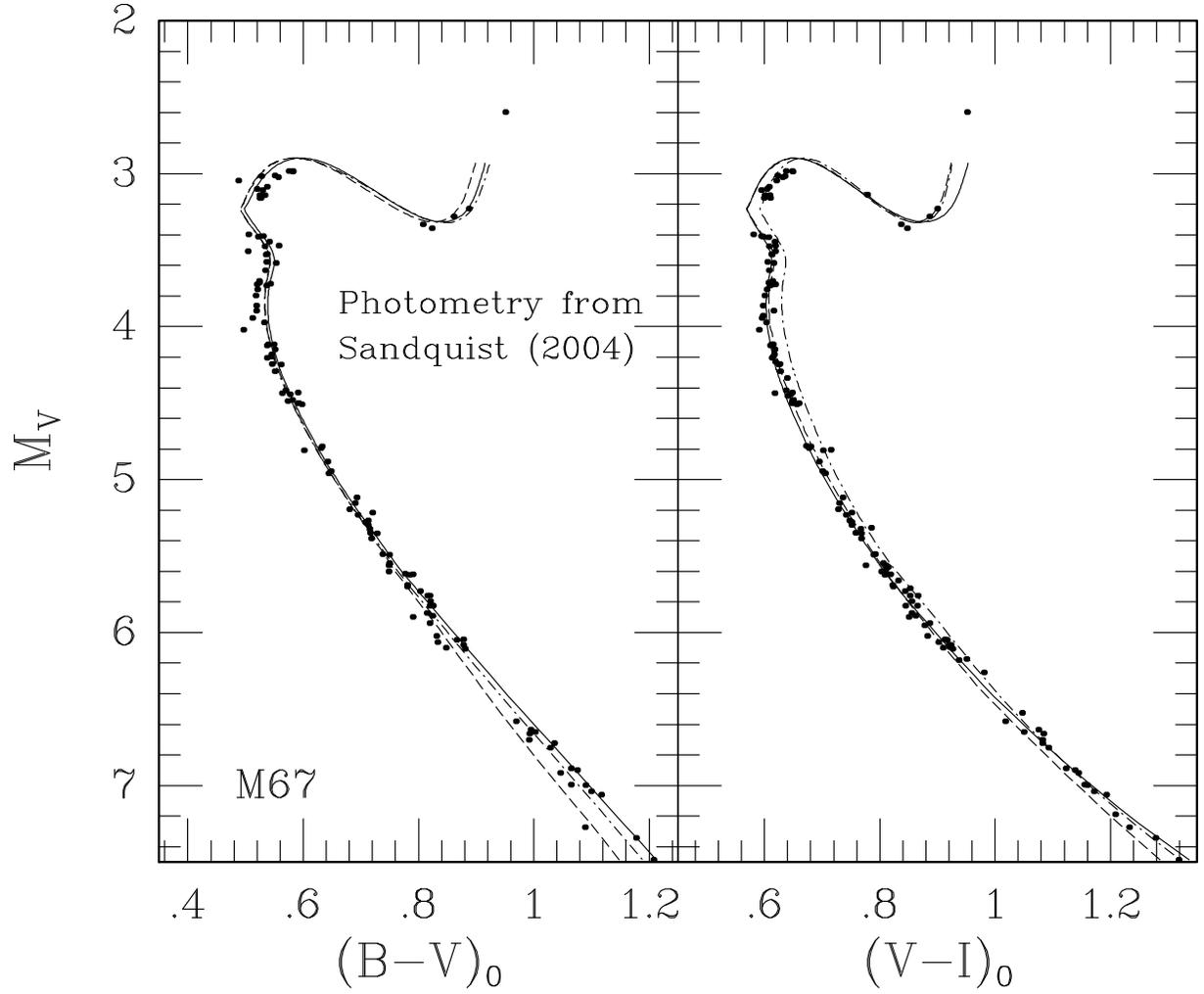}
\caption{Similar to the middle and left-hand panels of Fig.~\ref{fig:fig2}
except that the photometric observations are from \citet{san04}.}
\label{fig:fig3}
\end{figure}

Figure~\ref{fig:fig3} has been included in this study to show that the same
isochrones provide very similar fits to the CMDs of M$\,$67 reported by
\citet{san04}, which indicates that the Stetson and Sandquist data sets are in
close agreement.  This implies, in turn, that the (Cousins) $I$ magnitudes
and $V-I$ colors determined for the cluster MS and turnoff stars in most of the
CCD surveys of M$\,$67 prior to Sandquist's investigation (see \citealt{vs04})
are too bright/red by up to a few hundredths of a magnitude.  Such discrepancies 
are inherent to the VC03 $(V-I)$--$\teff$ relations for solar metallicity stars
because these transformations relied on the empirical constraints provided by
pre-2003 photometry of M$\,$67.  (In the case of solar-type stars having [Fe/H] 
$\approx 0.0$, the model $\teff$ scale is not a significant source of
uncertainty because it is precisely tied to the Sun through a calibration of
$\alpha_{\rm MLT}$.)  M$\,$67 thus provides a sobering example of how difficult
it is to obtain reliable CCD photometry to within 0.01 mag (\citealt{ste05}).

\begin{figure}
\plotone{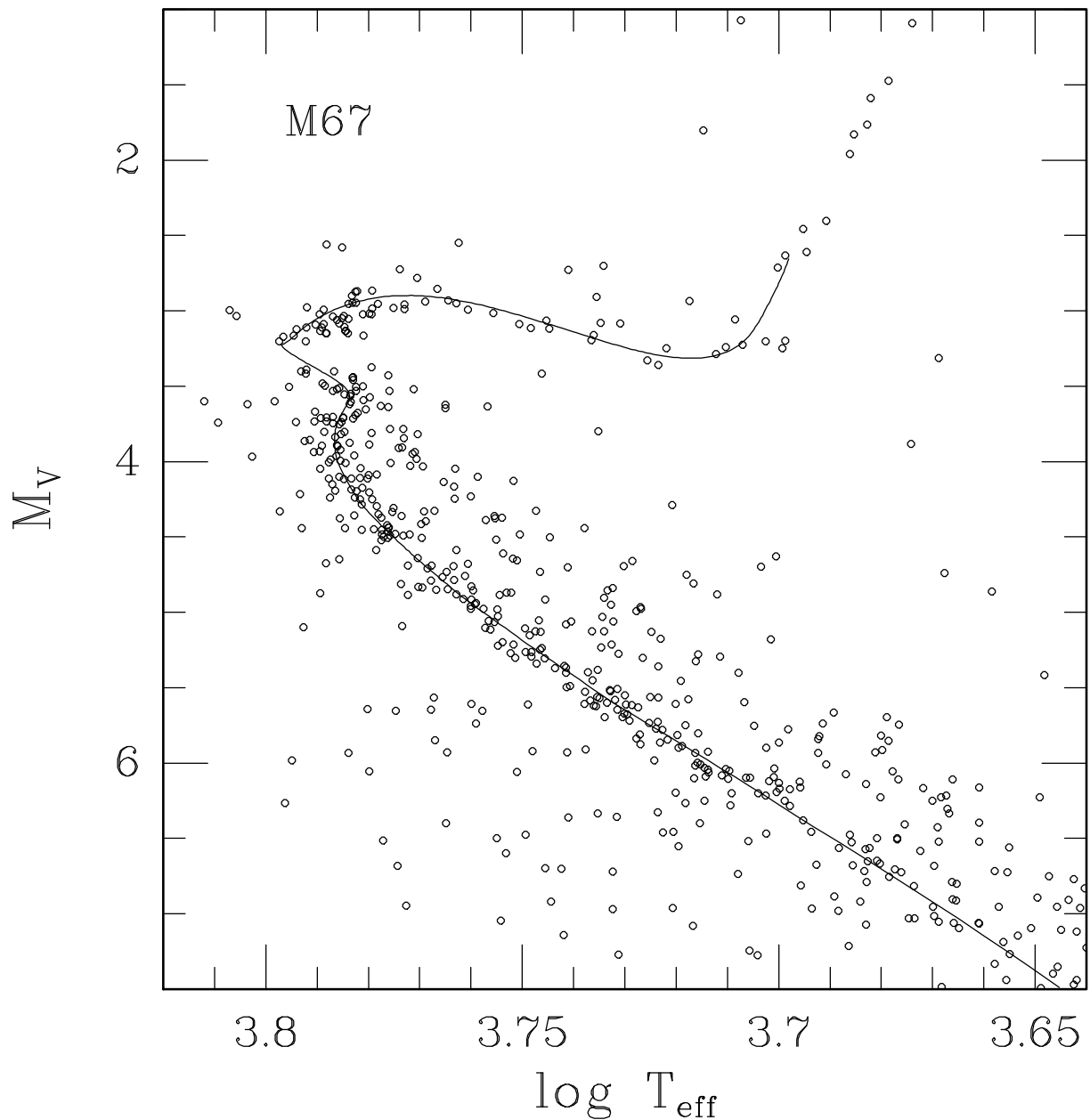}
\caption{Similar to Fig.~\ref{fig:fig2} except that the comparison is made on
the $(\log\teff,\,M_V)$-diagram.  The temperatures of M$\,$67 stars have been
derived from their $(V-K_S)_0$ indices using the CRMBA color transformations.}
\label{fig:fig4}
\end{figure}

The CRMBA and MARCS transformations to $V-K_S$ are evidently almost identical
for solar abundance stars (see the right-hand panel of Fig.~\ref{fig:fig2}), as
both result in in a nearly perfect superposition of the isochrone onto the 
observed CMD from at least $M_V = 7.5$ to the base of the RGB.  Indeed, if the
MARCS colors for low gravities were just $\sim 0.04$ mag redder, so that the
dashed curve lined up with the solid curve along the lower RGB, the models would
provide a very good match to the cluster giants as well.  A particularly
compelling illustration of the excellent consistency between theory and
observations is shown in Figure~\ref{fig:fig4}.  In this plot, the $(V-K_S)_0$
colors of M$\,$67 stars were converted to $\log\teff$ values using the CRMBA
transformations, which are valid only for dwarfs and subgiants, and then the
same isochrone that appears in the two previous figures was overlaid onto the
resultant $(\log\teff,\,M_V)$-diagram.  The solid curve reproduces the observed
morphology so well that it looks more like an estimate of the mean cluster
fiducial than a totally independent prediction from stellar evolutionary theory.
 
\subsection{NGC$\,$6791 ([Fe/H] $\approx +0.3$)}
\label{subsec:n6791}

It can be expected that it will be difficult to use NGC$\,$6791 to assess the
accuracy of the color--$\teff$ relations for super-metal-rich stars because even
an uncertainty of 0.1 dex in its [Fe/H] value, or in the m/H number abundance
ratios of many of the other metals, may affect some colors (notably $B-V$) more
than others (e.g., $V-K_S$).  The relatively high and uncertain foreground
reddening ($0.1\lta E(B-V) \lta 0.2$; see the discussion by \citealt{sbg03}) is
another complication, as is the cluster helium content, which is generally quite
hard to determine in old, metal-rich open clusters.  Fortunately, however,
eclipsing binaries have been found in this system, and the mass--radius diagram
derived from such stars can be used to determine the dependence of $Y$ on [Fe/H]
through comparisons with stellar models (\citealt{gch08}).  

In fact, a thorough study of NGC$\,$6791 and its binary stars is being carried
out by K.~Brogaard et al.~(2010, in preparation).  Very preliminary results
from this investigation suggest that, among other possibilities, the cluster
stars have $Y \approx 0.30$ if the adopted [Fe/H] value is $+0.30$
(\citealt{bjd09}), assuming that the metals are in the proportions given by
\citet{gs98}.  Moreover, an age near 8.0 Gyr is required to obtain consistent
fits to both the mass--radius and color--magnitude diagrams.  Since these
particular findings are based on models that were computed using the Victoria
stellar evolution code, we have chosen to compare just this one isochrone that
appears to provide a good fit to the observed properties of the cluster 
binaries with our $BVI$ photometry for NGC$\,$6791.

\begin{figure}
\plotone{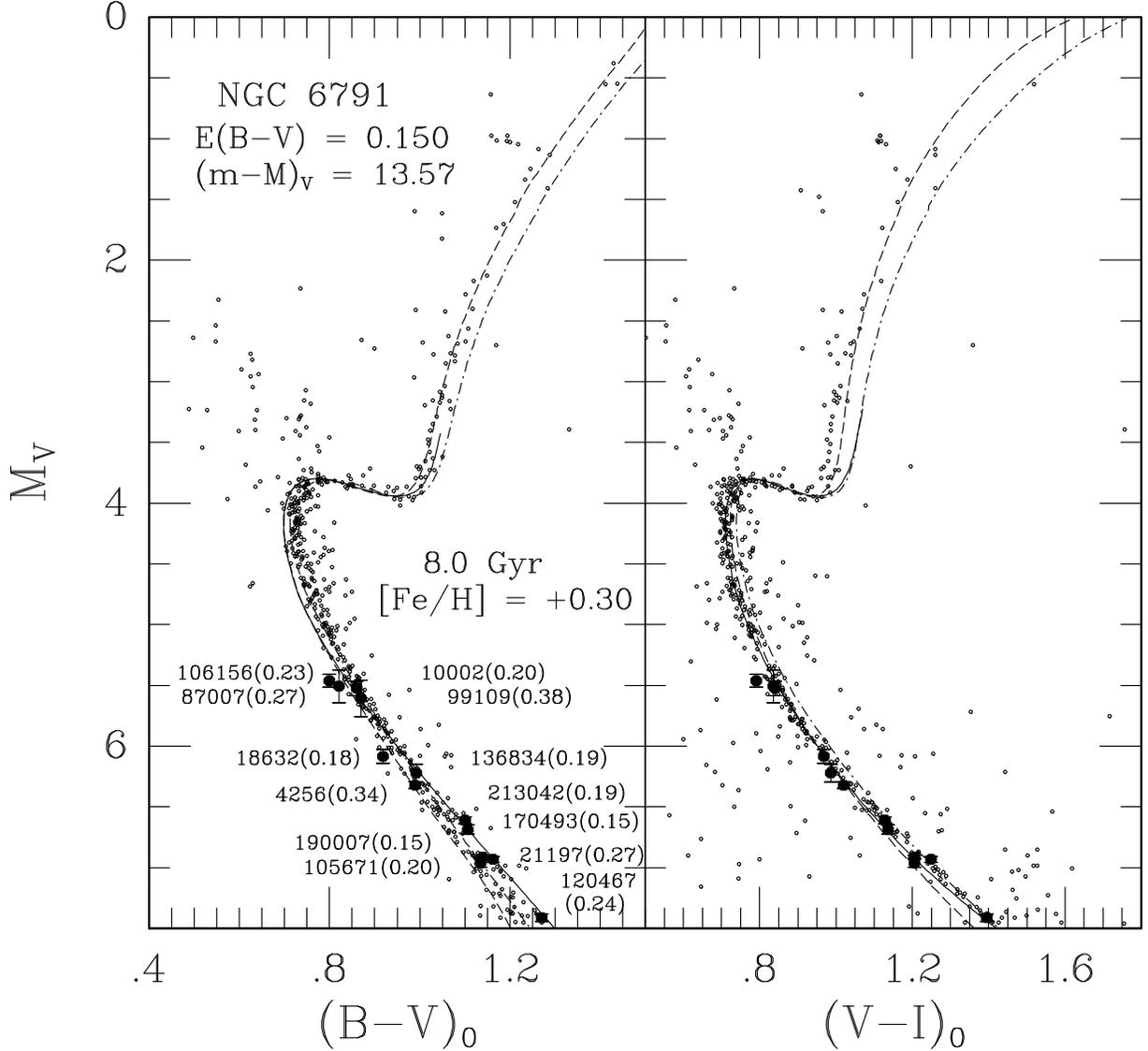}
\caption{Comparison of an 8.0 Gyr isochrone for [Fe/H] $=+0.30$ and $Y = 0.30$
with current $BV$ and $VI$ observations for NGC$\,$6791 from the on-going
\citet{ste00} project.  The adoption of the CRMBA, MARCS, and VC03 color
transformations are represented, in turn, by the solid, dashed, and dot-dashed
loci, respectively.  Filled circles represent those stars in the sample 
studied by CRMBA that have $0.15\le$ [Fe/H] $\le 0.45$ and $1\sigma$
uncertainties in their $M_V$ values of $\le 0.15$ mag based on {\it Hipparcos}
parallaxes (\citealt{vl07}).  The stars, which all have $E(B-V) = 0.0$
according to CRMBA, are identified by their ``HD" numbers: their [Fe/H] values
(from CRMBA) are given within the parentheses.  In the right-hand panel, the
observed turnoff is bluer than the values predicted by the CRMBA, MARCS, and
VC03 loci by about 0.005, 0.015, and 0.03 mag, respectively.  In the left-hand
panel, the observed turnoff is $\sim 0.03$ mag redder than the turnoffs of the
CRMBA- and MARCS-transformed isochrones, and $\approx 0.012$ mag redder than
the VC03-transformed isochrone.}
\label{fig:fig5}
\end{figure}

As shown in  Figure~\ref{fig:fig5}, this isochrone provides a rather good fit
to the cluster MS stars on the $[(V-I)_0,\,M_V]$-plane if $E(B-V) = 0.15$,
$E(V-I)/E(B-V) = 1.356$ (\citealt{mcc04}), $(m-M)_V = 13.57$, and either the
CRMBA or MARCS color transformations are assumed.  (The VC03 $V-I$ colors are
too red, for reasons that now appear to be understood; see \S~\ref{subsec:m67}.)
Given the likelihood that some fraction of the bluest stars at the top of the
main sequence are binaries, the isochrone faithfully reproduces the morphology
of the CMD from $\sim 2.5$ mag below the turnoff through to the lower RGB, where
the models seem to be $\approx 0.05$ mag too red.  The adopted reddening
agrees well with the determinations from the \citet{sfd98} dust maps, from the
properties of the cluster sdB stars (\citealt{kr95}), and from a fit of the
NGC$\,$6791 CMD to the {\it Hipparcos} CMD for solar neighborhood stars
(\citealt{slv03}).  Indeed, \citet{bsv10} have found that the same isochrone,
on the assumption of the same distance and $E(B-V)$ value, provides a fully
consistent fit to $VJK_S$ photometry of NGC$\,$6791.  To obtain such
consistency when the reddening corrections that are applied to the different
color indices vary so much --- since $E(V-J)/E(B-V) = 2.251$ and
$E(V-K_S)/E(B-V) = 2.72$ (\citealt{mcc04}) --- provides an especially strong
argument in support of the adopted reddening.

The fact that the CRMBA- and MARCS-transformed isochrones represent the $V-I$,
$V-J$, and $V-K_S$ observations of NGC$\,$6791 so well leads one to suspect that
inadequacies in the $(B-V)$--$\teff$ relations are responsible for the
discrepancies that are apparent in the left-hand panel of Fig.~\ref{fig:fig5}.
$B-V$ colors that are derived from model atmospheres and synthetic spectra are
bound to be more problematic than those photometric indices that involve redder
filter bandpasses than $B$, especially when the metallicity is high.  It is,
however, somewhat surprising that the slope of the solid curve is appreciably
shallower than that of the observed MS, given that no such problems are apparent
in any of the comparisons of isochrones with observations at longer wavelengths.

As a check of how well the CRMBA transformations reproduce the properties of
super-metal-rich stars, we have plotted in Fig.~\ref{fig:fig5} the dwarfs from
their paper having $0.15\le$ [Fe/H] $\le 0.45$, $M_V > 5.4$, and $\sigma_{M_V}
\le 0.15$ (based on parallaxes given in the New Hipparcos Catalogue by
\citealt{vl07}).  Although there are relatively few stars, they define quite a
tight sequence --- especially in the right-hand panel, where the field stars
provide a good match to the lower main-sequence of NGC$\,$6791, thereby offering
strong support for the adopted distance modulus (if the foreground reddening is
$E(B-V) \approx 0.15$).  The mean metallicity of the selected stars is [Fe/H] $=
0.23$ and, as it should, their mean locus on the $[(B-V)_0,\,M_V]$-diagram is
slightly to the blue of the solid curve, which assumes [Fe/H] $= 0.30$.  Thus,
the slope of the CRMBA-transformed isochrone in the left-hand panel of
Fig.~\ref{fig:fig5} is consistent with that implied by field dwarfs of high
metallicity.

Since the reddening of NGC$\,$6791 is fairly high, we also checked whether a
scaling of the reddening correction with the intrinsic color of the stars,
which is more correct than assuming a constant ratio between different bands
(see, e.g., Bessell et al.~1998), could introduce any significant differential
shift.  For the parameter space relevant to the MS of NGC$\,$6791, the 
differential effects between $E(B-V)$ and $E(V-I)$ amount to $\lta 0.01$ mag,
and the consequences for $M_V$ appear to be no more than $\sim 0.02$ mag.
Hence, the steeper MS in the left-hand panel of Fig.~\ref{fig:fig5} cannot be
explained in terms of reddening effects.  Perhaps chemical abundance
pecularities are responsible for the apparent difficulty of matching the $BV$
observations, or there are some systematic errors in the photometry, or maybe
the correct explanation is something entirely different.  Hopefully the
forthcoming paper by K.~Brogaard et al.~will shed some light on this issue.
As found for the other open clusters that have been considered, the VC03
$(B-V)$--$\teff$ relations produce the best overall match of the isochrone to
the $BV$ photometry for lower MS stars in NGC$\,$6791, though they are also
0.01--0.02 mag too blue in the vicinity of the turnoff. The MARCS
transformations appear to be the most realistic ones for super-metal-rich
giants.

\subsection{Field Subdwarfs ($-2.2\lta$ [Fe/H] $\lta -0.5$)}
\label{subsec:sbd}

Metal-poor dwarfs in the solar neighborhood provide stronger constraints on 
stellar models than their counterparts in globular clusters because many of the
former have well-determined distances (from {\it Hipparcos} observations) and
their temperatures and metallicities have been determined from high-resolution
spectroscopy.  Consequently, it makes sense to examine how well our isochrones
are able to reproduce the properties of the field subdwarfs before turning our
attention to GCs.  (Given the possibility of systematic errors in the derived
[Fe/H] values of field stars, which are often taken from different spectroscopic
studies, one might expect that the slope of the main sequence on the various
color-magnitude planes would be more reliably given by GCs.  However,
photometric calibrations may also suffer from such errors.)
 
Except for 4 stars, the sample of nearly 100 subdwarfs considered in this
section has been drawn from the paper by CRMBA (their Table~8).  We have opted
to select those stars for which $BV(RI)_C$ photometry is given that have
$\sigma_\pi/\pi\lta 0.15$ and metallicities in the 
range $-2.2\lta$ [Fe/H] $\lta -0.5$.  A few stars that did not satisfy these
criteria were included either to augment the number of stars with [Fe/H] $\lta
-1.5$ or to increase the number in common to the studies by CRMBA and R.~Gratton
and collaborators (hereafter referred to as the ``Gratton" sample --- from
Gratton et al.~1996, \citealt{cge99}, or by private communication; for details,
see \citealt{bv01}).  Four additional subdwarfs, not considered by CRMBA, but
which satisfied the above constraints, were added to the sample: their basic
parameters were derived by Clementini et al.  As far as their distribution with
metal abundance is concerned, the CRMBA sample is composed of 39, 25, 21, and 8
stars in the four 0.5 dex intervals of [Fe/H] between $-0.50$ and $-2.50$, in
turn, whereas the Gratton sample consists of 18, 13, 13, and 4 stars in the
same metallicity bins.

It has already been mentioned (see \S~\ref{sec:intro}) that isochrones provide
a good fit to the classic Population II subdwarfs on the
$(\log\teff,\,M_V)$-diagram {\it if} their temperatures are close to those
determined by Gratton et al.  (We refer here to the 6--10 stars that have been
used for many years to derive the distances to GCs via the main-sequence-fitting
technique; see, e.g., \citealt{rfv88}; \citealt{sbs96}.)  As shown by VandenBerg
(2008; see his Fig.~1), these temperature estimates are $\sim 75$--100~K warmer,
in the mean, than those derived by, e.g., \citet{aam96}; \citet{msv06}; and
\citet{cps07}.  In fact, CRMBA favor even higher temperatures.

\begin{figure}
\plotone{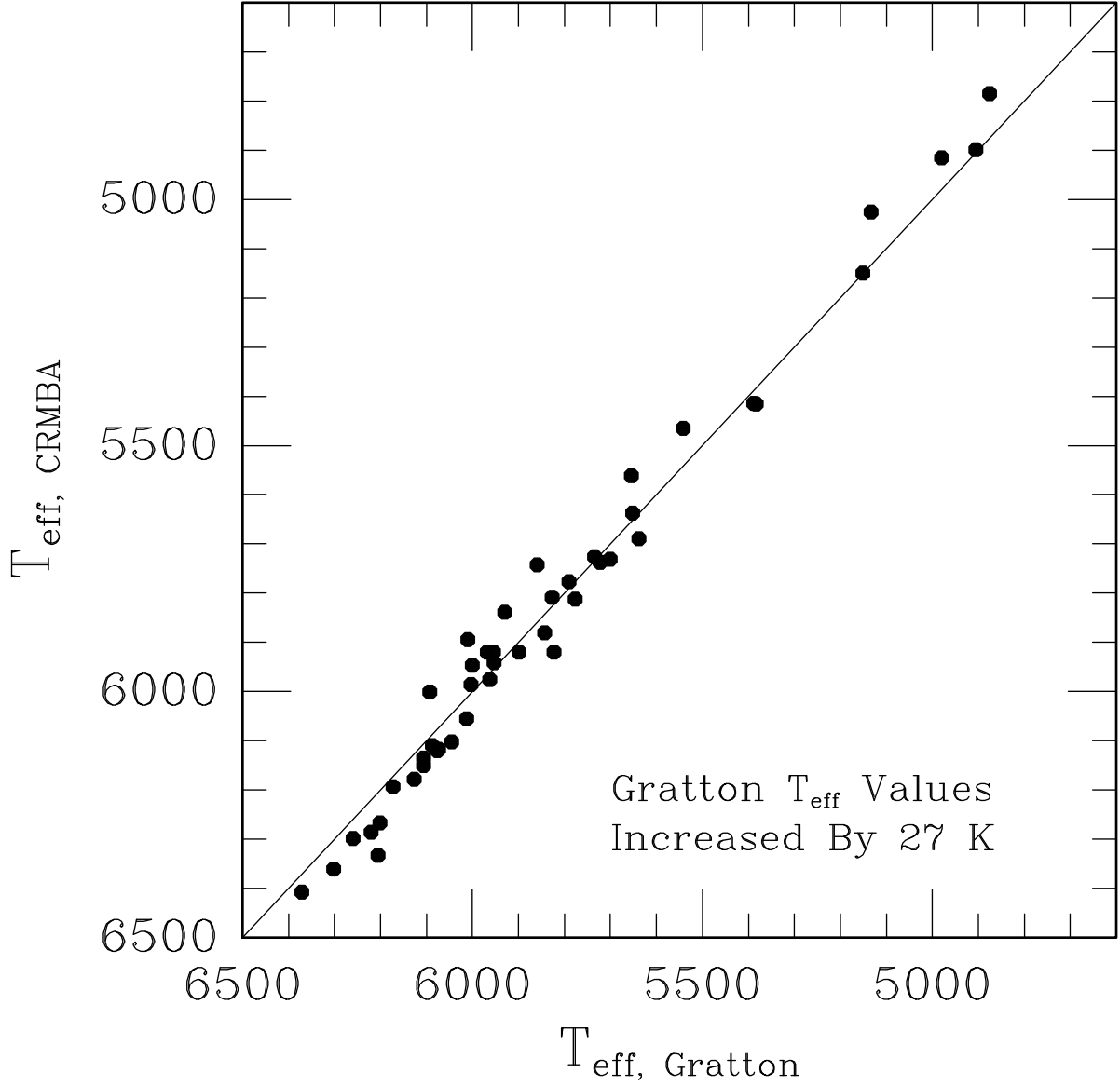}
\caption{Comparison of the effective temperatures derived by CRMBA and by
R.~Gratton and colleagues (see the text) after the latter values were increased
by 27 K (as noted), which is the mean difference between the two data sets.}
\label{fig:fig6}
\end{figure}

\begin{figure}
\plotone{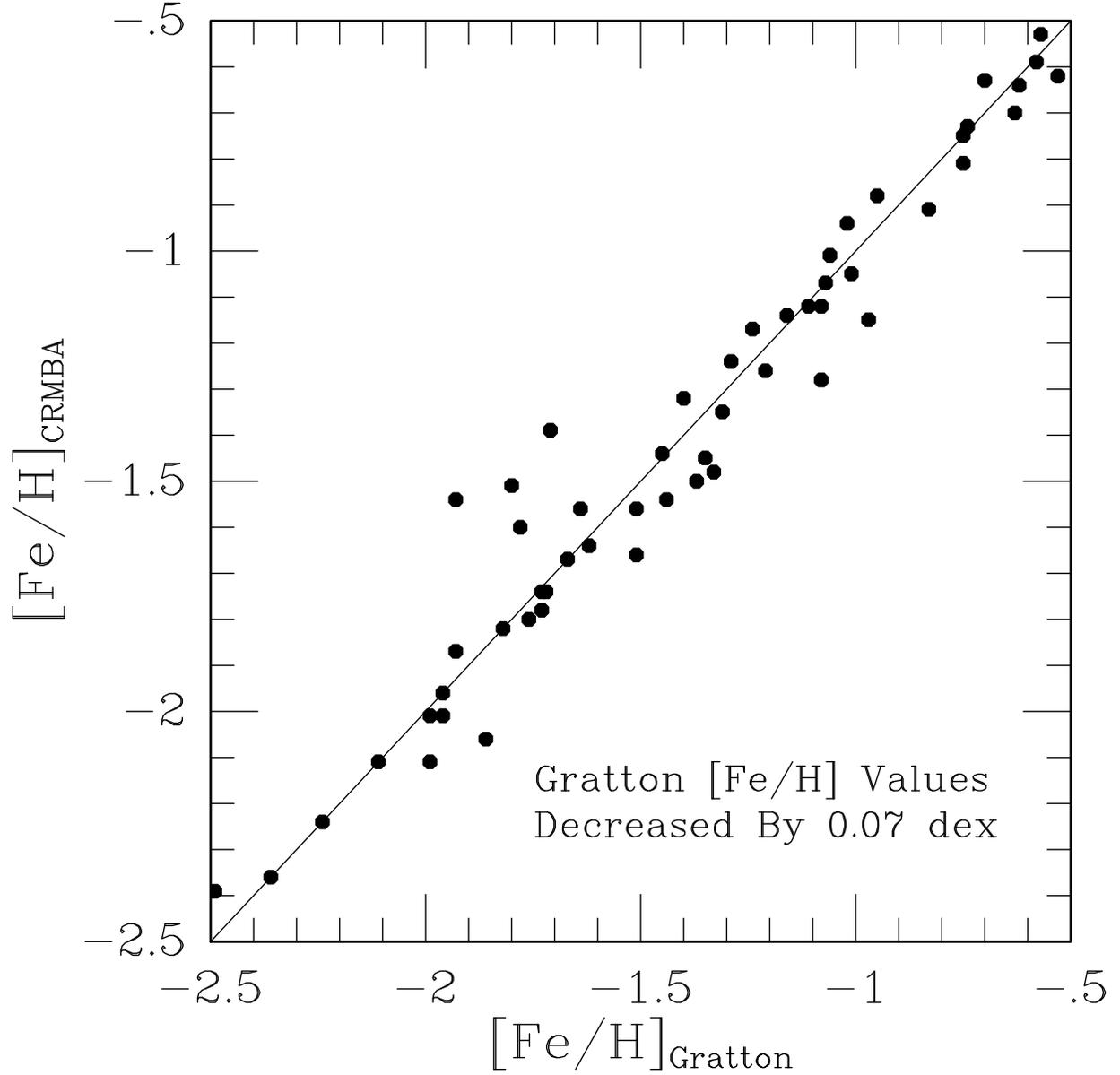}
\caption{As in the previous figure, except that the [Fe/H] values determined
by the two groups are compared (after the indicated adjustment to the Gratton
iron abundances was made).  Note that the sample of stars considered here was
restricted to [Fe/H] values from $-2.5$ to $-0.5$.}
\label{fig:fig7}
\end{figure}

Figure~\ref{fig:fig6} compares the $\teff$ values given by CRMBA and Gratton et
al.~for the 45 stars in our sample that were studied by both groups.  The
Gratton temperatures are cooler than those of CRMBA by 27 K, on average, so the
former values were increased by this amount prior to being plotted.  There is
clearly some systematic variation in the scatter of the points about the
diagonal ``line of equality", which indicates that the temperature differences
actually range from near 0 K for the coolest stars to $\approx 70$ K for the
warmest ones.  Curiously, the [Fe/H] values adopted by CRMBA are an average of
0.07 dex {\it less} than the Gratton determinations (see Figure~\ref{fig:fig7}),
despite the expectation that the warmer temperatures of CRMBA would require
higher, not lower, metal abundances in order to match the observed strengths of
spectral lines.  However, the [Fe/H] values given by CRMBA were collected from
the literature and, although they are accurate enough for the purpose of their
IRFM calculations, which depend only mildly on metallicity, it is not surprising
to find the aforementioned differences.  {\it Independent} determinations of a
star's metal abundance can easily, and often do, vary by at least 0.1 dex,
which is typically taken to be the uncertainty of such measurements.  In any
case, such small differences in [Fe/H] have only minor effects on the colors
that are derived from color--$\teff$ relations, which are mainly a function of
temperature (especially with decreasing metallicity).

Since the CRMBA transformations are based on their estimates of the basic
properties of the same stars considered here (along with many more), they will
necessarily yield color indices that, in the mean, agree very well with those
observed.  How well, then, will the MARCS and VC03 color--$\teff$ relations be
able to reproduce the observed subdwarf colors if the assumed temperatures,
gravities, and [Fe/H] values are as given by CRMBA?  The answer to this
question is given in Figure~\ref{fig:fig8}.  This shows that the predicted
$B-V$ indices are too blue by $\sim 0.02$--0.03 mag (as noted in the uppermost
panels), while the offsets in the $V-R$ and $V-I$ colors are $\lta 0.01$ mag
(see the middle and bottom row of panels).  Indeed, the model-atmosphere-based
$B-V$ colors, in particular, tend to become more discrepant for redder stars.
(It is somewhat disconcerting to find that the MARCS and VC03 transformations
to $B-V$ appear to have more difficulty reproducing the observed colors of
metal-poor stars, even relatively warm ones, than of those having close to the 
solar metallicity: recall our analyses of the M$\,$67 and Hyades CMDs.
Nevertheless, this is apparently the case.)

If, however, we consider the Gratton sample instead --- i.e., we interpolate
in the MARCS and VC03 transformations to derive colors on the assumption of the
$\teff$, $\log\,g$, and [Fe/H] values derived by Gratton et al.~--- the results
are quite different.  As shown in Figure~\ref{fig:fig9}, the predicted $B-V$
indices are now able to reproduce the observed $B-V$ colors quite well, but only
at the cost of worsening the agreement in the case of $V-R$ and (especially)
$V-I$.  [Assuming lower temperatures than those given by Gratton et al.~by
$\approx 100$ K, so as to be closer to the Alonso et al.~(1996) and Cenarro
et al.~(1997) $\teff$ scales, would result in $B-V$, $V-R$, and $V-I$ colors
from the MARCS transformations that are an average of 0.017, 0.022, and 0.040
mag too red, respectively.  This shows that the relatively low temperatures
in such studies as Alonso et al.~and Cenarro et al.~are not consistent with the
photometric $\teff$ scale implied by the CRMBA IRFM zero-points and absolute
calibration, together with the MARCS model atmospheres.]  If the predicted
$V-R$ and $V-I$ colors are more trustworthy than $B-V$, we would conclude that
Fig.~\ref{fig:fig8} is closer to the truth than Fig.~\ref{fig:fig9}; i.e., the
CRMBA $\teff$ scale is more realistic than the one derived by Gratton et al.
As there is already reasonably good agreement between the CRMBA, MARCS, and (to
a lesser extent) VC03 $(V-R)$--$\teff$ and $(V-I)$--$\teff$ relations, very
good consistency of those for $B-V$ can be obtained as well if the MARCS and
VC03 transformations to $B-V$ were adjusted to the red by $\approx 0.02$--0.03
mag.

The next step in our analysis is to determine how well the empirically derived
temperatures of the subdwarfs agree with those predicted by stellar models.
Usually (see, e.g., \citealt{van08}) this involves the construction of a
so-called ``mono-metallicity" subdwarf sequence, whereby isochrones are used
differentially to correct the measured $\teff$ of each subdwarf to the
temperature it would have, at its observed $M_V$, if it had a particular
(reference) [Fe/H] value.  Once such adjustments have been made to all of the
subdwarfs, thereby compensating for metallicity differences, the latter are
superimposed on an isochrone for the reference metallicity and some assessment
made of the level of agreement between the two.  To ensure that the results are
essentially independent of the age of the isochrone, subdwarfs brighter than,
say, $M_V = 5.0$ are generally excluded from such comparisons.

A different approach is adopted here.  To be specific, those subdwarfs with 
well-determined $M_V$ values are superimposed directly onto a set of isochrones
for a fixed age (12 Gyr) and helium abundance ($Y = 0.25$) and a range in [Fe/H]
from $-2.4$ to $-0.6$ (with [$\alpha$/Fe] $\approx 0.4$), when plotted on the
$(\log\teff,\,M_V)$-diagram (see the bottom panel of Figure~\ref{fig:fig10}).
The extent to which isochrones for the {\it observed} metallicities are able to
reproduce the subdwarf luminosities and temperatures clearly provides a powerful
test of the models in an absolute sense.  As there is a large number of
metal-rich stars in the CRMBA sample with precise parallaxes
(from \citealt{vl07}), we have opted to use only those subdwarfs with
$-1.0\le$ [Fe/H] $\le -0.5$ and $M_V \ge 5.2$ that have $\sigma_{M_V} \le 0.1$.
The majority of the more metal-deficient stars considered here also satisfy
these constraints, though any such star with $M_V \ge 4.4$ and $\sigma_{M_V}
\le 0.15$ was included because of the paucity of good subdwarf standards with
[Fe/H] $< -1.0$.  No star that fulfilled these criteria was rejected, unless
it is known to be a binary, even though some of them (e.g., BD$\,+$41$\,$3306,
HD$\,$145417) appear to have anomalous locations on the H-R diagram relative
to those of most of the stars that have similar metal abundances.  The
resultant data set consists of 33 stars, of which 11 have [Fe/H] $< -1.2$ (the
filled circles in Fig.~\ref{fig:fig10}), and the rest are more metal rich (the
open circles).

To demonstrate that age uncertainties do not play a significant role in this
comparison of theory with observations, 10 Gyr isochrones for [Fe/H] $=-2.40$,
$-1.40$, and $-0.60$ have also been plotted (as dashed curves).  Only at 
$M_V \lta 5.2$ are the differences in $\teff$ given by isochrones that differ
in age by 2 Gyr larger than $\sim 30$~K.  Even though a few of the subdwarfs
are brighter than this absolute magnitude, their ages are probably closer to
12 Gyr than to 10 Gyr, if they are coeval with GCs having similar metallicities
(see the fits of isochrones to GC CMDs presented below).  Consequently, the
differences between the solid and dashed curves may be an overestimate of
the actual uncertainties in the isochrone locations due to age.  (As most of
the subdwarfs are fainter than $M_V = 5.2$, age uncertainties are clearly of
no concern for them.)     

At the observed $M_V$ of each subdwarf, the isochrones define a relationship
between $\log\teff$ and [Fe/H] that can be readily interpolated (or
extrapolated) to yield the metal abundance corresponding to the temperature
given by CRMBA for that star.  The difference between the observed [Fe/H] and
that implied by the isochrones is plotted as a function of $\log\teff$ in the
middle panel, which indicates that there is rather good consistency between
the observations and the isochrones.  As noted in this panel, and indicated by
the arrow, the mean value of $\delta$[Fe/H] is $-0.03$, in the sense that the
subdwarfs are slightly more metal-poor than one would infer from their locations
relative to the isochrones (in the lower panel).  There is no obvious trend in
the $\delta$[Fe/H] values with temperature, except in the case of the coolest
stars, though unrecognized binaries may be responsible for some fraction of 
asymmetry.  It has long been a puzzle, for instance, why HD$\,$145417 and
HD$\,$25329 lie above the mono-metallicity subdwarf defined by HD$\,$64090,
HD$\,$103095, HD$\,$134439 and HD$\,$134440 (e.g., see \citealt{bv01}).  Only
one star (HD$\,$193901) has a value of $\delta$[Fe/H] larger than $\pm 0.6$;
hence, the metallicity given by CRMBA for this star is quite inconsistent with
its location on the H-R diagram.

One can also interpolate within the isochrones to determine the correction to
the temperature of each subdwarf that would be required to place it on the
isochrone which has the same metal abundance as the subdwarf.  The differences
in $\teff$ so derived are plotted in the upper panel of Fig.~\ref{fig:fig10} as
a function of $\log\teff$.  Here, as well, the majority of the points lie
slightly below the dashed line, which indicates that the temperatures given by
CRMBA for these stars are less than those implied by the isochrones: the mean
value of $\delta\teff$ is just $-8$~K (as noted within the panel and indicated
by the arrow).  A few stars, including BD$\,$+41$\,$3306, HD$\,$193901,
HD$\,$145417, and HD$\,$25329, have $\delta\teff$ offsets that are larger than
$\pm 100$~K.\footnote{It is easy to identify the most discrepant points, should
anyone wish to do so.  Since the same abscissa applies to all three panels, a
vertical line through a star's position in the bottom panel will pass through
the points representing that star in the uppermost panels if its $\delta$[Fe/H]
and/or $\delta\teff$ values are within the ranges plotted.}  Nevertheless, the
temperature scale predicted by the isochrones is clearly very similar to that
favored by CRMBA.  (Even though there is a tendency for the models to be
somewhat too warm relative to the observations of the faintest stars, it must
be kept in mind that the comparisons presented in Fig.~\ref{fig:fig10} are
{\it very} sensitive to the adopted [Fe/H] values.  If the metallicities of
the faintest, and hence coolest, stars have been underestimated by as little as
0.15 dex, the temperatures inferred for them from the isochrones would be much
more consistent with those tabulated by CRMBA.  To be sure, it is also possible
that the systematic errors in the model $\teff$ scale are responsible for the
noted tendency.)

(As far as we have been able to determine, Gratton et al.~have not studied 
most of the subdwarfs that are identified in Fig.~\ref{fig:fig10}.  They do
provide metallicities and temperatures for all of stars that have been plotted
as filled circles, except HD$\,$97320, as well as for HD$\,$201891.  A similar
analysis of just those 11 stars --- not shown here --- reveals that their
estimates of [Fe/H] are 0.02 dex more metal rich and their $\teff$ values warmer
by 6 K, in the mean, than those implied by the isochrones.  Because the Gratton
sample is so small, the remainder of this section will consider only the
subdwarfs plotted in Fig.~\ref{fig:fig10} for which CRMBA provide $BV(RI)_C$
photometry as well as their estimates of $\teff$, $\log\,g$, and [Fe/H].) 

Figure~\ref{fig:fig11} repeats the comparisons shown in the previous figure,
except that the isochrones have been transformed to the three color planes using
the MARCS color--$\teff$ relations.  For each subdwarf, the [Fe/H] value given
by CRMBA minus the [Fe/H] value of the isochrone that intersects its location on
the CMD is plotted in the middle panel, whereas the uppermost panel plots the
observed subdwarf color minus that of the isochrone for the star's metallicity
(at the same absolute magnitude).  It is quite clear, for instance, that the
isochrones are too blue on the $[(B-V)_0,\,M_V]$-diagram, since most of the 
points have negative $\delta$[Fe/H] values and positive $\delta$(color) values.
The mean offsets are $-0.18$ dex and 0.031 mag, respectively.  That is, if the
isochrone $B-V$ colors were adjusted redward by 0.031 mag, the mean residuals
would be zero and the consistency between theory and observations would be
about as good as one could possibly get without culling stars from the sample.

Because both $V-R$ and $V-I$ are much less sensitive to metallicity than $B-V$,
especially at low [Fe/H] values (compare the bottom three panels of
Fig.~\ref{fig:fig11}), moderately large values of $\delta$[Fe/H] translate to
small values of $\delta$(color), as shown in the respective middle and uppermost
panels.  As a consequence, the apparent trends of the $\delta$[Fe/H] values
with $V-R$ and $V-I$ are misleading, except in the case of the reddest stars,
which are discrepant in the same sense as found in the previous figure.  They
simply reflect the fact that the upper MS segments of the isochrones for
metal-poor stars are close together.  Indeed, the upper panels show that,
bluer than $(V-R)_0 = 0.55$ or $(V-I)_0 = 1.0$, the stars are all quite close
to the dashed lines, and hence that there is excellent consistency between the
predicted and the observed colors (i.e., they differ by $\lta 0.02$ mag).
Insofar as the reddest stars are concerned, it seems more likely that
the discrepancies are indicative of a problem with the subdwarf [Fe/H] values
than with the model temperatures because the same isochrones provide excellent
fits to lower-main-sequence observations of GC stars on the
$[(V-I)_0,\,M_V]$-diagram to at least $(V-I)_0 = 1.20$ (see the plots presented
in the following sections).

The consequences of using the VC03 color--$\teff$ relations instead of the
MARCS transformations are shown in Fig.~\ref{fig:fig12}.  Although the former
predict somewhat redder $B-V$ colors than the latter, the observed colors of
the subdwarfs are still 0.018 mag redder, on average, than those inferred from
the isochrones, reflecting the fact that the observed and predicted [Fe/H]
values differ by the mean value of $-0.13$ dex.  Interestingly, there is little
to distinguish the $\delta$[Fe/H] and $\delta(V-R)$ plots when either the MARCS
or VC03 transformations are assumed, but just as was found in our consideration
of metal-rich open clusters (see \S~\ref{subsec:hyades} and~\ref{subsec:m67}),
the subdwarfs indicate that the VC03 transformations to $V-I$ are too red by
about 0.02 mag.  It is also worth noting that the open circles, which represent
stars having [Fe/H] $> -1.2$, exhibit much more scatter than the filled
circles, which represent lower metallicity stars.  Metal abundance uncertainties
will translate into a larger scatter at higher [Fe/H] values simply because the
dependence of $V-I$ on [Fe/H] increases as the metallicity increases.
 
Figure~\ref{fig:fig13} shows that VR2010 isochrones, together with the CRMBA
empirical color--$\teff$ relations, provide the best overall match to the local
subdwarfs on the three color planes considered in this investigation.  For the
$B-V$, $V-R$, and $V-I$ panels, in turn, the mean values of $\delta$[Fe/H] are
$-0.04$, $+0.10$, and $+0.04$ dex, whereas the mean values of $\delta$(color)
are $+0.007$, $-0.003$, and $-0.004$ mag.  Such small differences in the mean
offsets indicate that there is very good consistency between the models and the
best-observed subdwarf standards if their properties (i.e., temperatures,
gravities, and metallicities) are close to those adopted by CRMBA.  The
$V-R$ and $V-I$ observations of the most metal-deficient subdwarfs, in 
particular (i.e., the filled circles), are especially well reproduced by the
isochrones.  It is also worth pointing out that some stars, which appeared
anomalous in the $[\log\teff,\,M_V]$-diagram presented in Fig.~\ref{fig:fig10}
(notably BD$\,+$41$\,$3306 and HD$\,$25329) are well matched by the isochrones
for the observed metallicities on the $V-R$ and $V-I$ color planes, in
particular (see Fig.~\ref{fig:fig13}).  The latter also shows that some stars
(e.g., HD$\,$144579, HD$\,$216777) are matched by isochrones having quite
different [Fe/H] values on different CMDs.  It would be worth the time and
effort to study such stars (indeed, all of the subdwarfs) further to improve
our understanding of these very important calibrators of GC distances.  We
now turn to a consideration of a small number of GCs with [Fe/H] values 
spanning the range in [Fe/H] from $-0.8$ (47 Tucanae) to $-2.4$ (M$\,$92).   

\subsection{47 Tucanae ([Fe/H] $\approx -0.8$)}
\label{subsec:47tuc}

Using an innovative statistical analysis of new, extensive photometry for
47 Tucanae, \citet{bs09} have produced particularly tight and well-defined
fiducial sequences for this GC on the $B-V,\,V$ and $V-I,\,$V planes.  They
showed that Victoria-Regina isochrones (VandenBerg et al.~2006) for [Fe/H]
$= -0.83$ and [$\alpha$/Fe] $=0.3$ reproduced these sequences very well from
several magnitudes below the turnoff to the RGB tip if $E(B-V) = 0.04$ and
$(m-M)_V = 13.375$.  The assumed distance modulus was obtained from a fit of a
theoretical zero-age horizontal branch (ZAHB) locus for the aforementioned
metallicity to the lower bound of the distribution of cluster HB stars, as
well as from a fit of the cluster MS to local subdwarfs having a similar metal
abundance.  Guided by these results, we have chosen to compare VR2010 isochrones 
for [Fe/H] $=-0.80$, which is close to the latest estimates from high-resolution
spectroscopy (\citealt{km08}; \citealt{cbg09}), and [$\alpha$/Fe] $\approx 0.4$
to the Bergbusch-Stetson CMDs. (The assumed enhancements in the abundances of
the individual $\alpha$-elements range from 0.25 to 0.50, with a mean value of
about 0.4; see the VR2010 paper.)

As shown in Figure~\ref{fig:fig14}, fits of the cluster fiducials to the 
CRMBA-transformed isochrones for these abundances at $M_V \gta 6$ yield
$(m-M)_V \approx 13.40$ if $E(B-V) = 0.032$ (Schlegel et al.~1998) and
$E(V-I)/E(B-V) = 1.356$ (\citealt{mcc04}).  (This is equivalent to performing
main-sequence fits to the local subdwarfs, in view of the fact that the CMD
locations of such stars are quite well represented by isochrones that employ
the CRMBA color--$\teff$ relations; as shown in the previous section.)  The
apparent distance modulus which is derived in this way is clearly a compromise
since the observations are offset from the model loci in different directions in
the two panels: the solid curve lies along the blue edge of the 47 Tuc fiducial
on the $[(V-I)_0,\,M_V]$-plane, while it coincides with the red edge of the MS
observations in the $[(B-V)_0,\,M_V]$-diagram.  In order for the models to
match the luminosity of the cluster subgiant branch, an age near 11 Gyr is
required.  (Our age estimate is 1 Gyr less than the age inferred from the same
data set by Bergbusch \& Stetson.  We note that, in addition to minor
differences in the adopted chemical abundances and the derived distance, only
the isochrones used in the present study take the gravitational settling of
helium into account.  This has the effect of reducing the age at a given
turnoff luminosity by $\sim 8$--10\%.)

The largest discrepancies between theory and observations occur along the RGB
in the left-hand panel of Fig.~\ref{fig:fig14}, which suggests that the 
model-atmosphere-based transformations to $B-V$ for low-gravity stars having
[Fe/H] $\approx -0.80$ are up to $\sim 0.1$ mag too blue (though this is not
necessarily the correct explanation; see the next section).  If a higher metal
abundance were assumed, the corresponding isochrones would undoubtedly provide
a better match to the observed RGB on the $[(B-V)_0,\,M_V]$-diagram, but at
the expense of worsening the fit of the models to the $VI$ photometry.  The
main mismatch in the right-hand panel occurs near the turnoff, where the
isochrones are too blue independently of the color transformations that are
used.  We are not able to provide a good explanation for this problem, as the
photometric calibrations appear to be robust, and anything that alters the model
$\teff$ scale would work in opposite directions in the two color planes.

It is not impossible that the CRMBA transformations to $B-V$ for [Fe/H]
$\approx -0.80$ and temperatures appropriate to the MS and turnoff stars of
47 Tuc are too red by $\sim 0.02$ mag.  Indeed, the comparison of the
CRMBA-transformed isochrones to the local subdwarfs with [Fe/H] $\ge -1.2$ (the
open circles in the left-hand panels of Fig.~\ref{fig:fig13}) would be improved
if the model colors were adjusted to the blue by $\approx 0.02$ mag.  Such an
adjustment would also result in better consistency of the fits to the $BV$ and
$VI$ observations, even though some difficulties would persist in an absolute
sense.  (In this and some of the other figures contained in this paper, there
is a tendency for the CRMBA-transformed isochrones to be too red just at the
base of the giant branch.  This is possibly a consequence of the fact that
there is no explicit gravity dependence in these color--$\teff$ relations
and/or reflect the tendency of the adopted functional form to diverge at the
coolest temperatures.  Further work is clearly needed to resolve these issues.)
We note, finally, that isochrones using the VC03 semi-empirical transformations
provide a better fit to the {\it slope} of the lower MS on both color planes
than isochrones that employ the MARCS transformations --- as found in our
consideration of metal-rich open clusters as well. 

\subsection{NGC$\,$1851 and M$\,$5 ([Fe/H] $\approx -1.4$)}
\label{subsec:n1851}  

Unfortunately, NGC$\,$1851 was not one of the 19 globular clusters that was used
by \citet{cbg09} to calibrate their new metallicity scale, nor was it among the
sample that defined the original \citet{cg97} scale.  Although the former 
provide updated metallicities for most of the Galactic GCs, by performing 
weighted averages of data from different sources, their estimate of the metal
abundance of NGC$\,$1851, which is [Fe/H] $=-1.18$, may be too high by $\sim
0.2$ dex --- notwithstanding the fact that a similar value was obtained by
\citet{ki03} from their spectroscopic analysis of Fe II lines.  According to
\citet{zw84}, whose metallicity scale is still widely used, NGC$\,$1851 has
[Fe/H] $=-1.36$, which is only 0.04 dex greater than their value for M$\,$5.
Interestingly, the Carretta-Gratton determination of $-1.11$ for M$\,$5 has
been revised to $-1.35$ in the 2009 study by Carretta et al.~(who used M$\,$5
as one of the primary calibrating clusters).  Also worth noting is the fact
that stellar models have tended to favor [Fe/H] $\approx -1.4$ for both M$\,$5
and NGC$\,$1851 (e.g., \citealt{van00}) for the following reason.

If the $E(B-V)$ values given by the Schlegel et al.~(1998) dust maps are
assumed, and the dereddened CMDs for these two clusters are overlaid in such a
way that their red horizontal branch populations have the same luminosities,
then their main sequences superimpose on one another almost perfectly.  What
does differ, as discussed recently by \citet{ste09}, is the location of their
giant branches: the RGB of NGC$\,$1851 is significantly redder than that of
M$\,$5 at the same $M_V$ (on all CMDs).  The cause of this color shift is not
known, though differences in the [m/Fe] ratios of some elements could well be
the explanation given that recent evolutionary computations (see
\citealt{dcf07}; VR2010) have shown that variations in the abundances of some
metals (notably Mg and Si) will have important consequences for the
temperatures, and therefore the colors, of giant stars, even at low [Fe/H]
values.

What is important for the present investigation is that (i) if $E(B-V) = 0.034$
(Schlegel et al.~1998) and $(m-M)_V \approx 15.50$, which is obtained from MS
fits of the cluster fiducials to the CRMBA-transformed isochrones (and hence to
the subdwarf standards), then both the $BV$ and $VI$ observations {\it for the
lower main-sequence stars} in NGC$\,$1851 can be matched quite well by
theoretical isochrones for [Fe/H] $= -1.40$ and [$\alpha$/Fe] $\approx 0.4$ (see
Figure~\ref{fig:fig15}), and (ii) the same isochrones provide an equally good
match to the MS of M$\,$5 if its reddening and distance are similarly derived
(see Figure~\ref{fig:fig16}).  Indeed, as already mentioned, this is to be
expected if the assumed distance moduli are such that the red HB populations
in the two GCs are made coincident.  If the [Fe/H] values of NGC$\,$1851 and
M$\,$5 truly do differ by $\approx 0.2$ dex (and if their reddenings have been
accurately determined), then it must be the case that variations in the
abundances of some other elements compensate for the difference in iron content
in such a way as to produce CMDs for their MS stars that are nearly identical.
(VR2010 models will be used to explore this possibility.)

As our $R$ photometry for NGC$\,$1851 is not of the same quality as $B$, $V$, 
and $I$, little can be said about the predicted $V-R$ colors for MS stars except
that they appear to be too red by $\sim 0.03$ mag in the vicinity of the
turnoff.  The main point that can be made about the $(V-R)$--$\teff$ relations
is that the cluster giants are consistently fitted by the isochrones on both
the $V-R$ and $V-I$ color planes.  Fortunately, the fiducial sequence for dwarf
stars is much better defined in M$\,$5 than in NGC$\,$1851, and the middle panel
of Fig.~\ref{fig:fig16} indicates that the isochrones actually provide a good
fit to the observations at $M_V \gta 5$.  To have a completely consistent
interpretation of the data at $3\lta M_V \lta 5$ on all three color planes, the
isochrone $V-R$ colors should be corrected to the blue by $\approx 0.02$ mag, 
or less, if some fraction of this offset can be attributed to the calibration
of the photometry.  In fact, the observed $V-R$ colors have an uncertainty of
at least $\pm 0.01$ mag.  Because the same isochrones reproduce the entire CMD
of M$\,$5 on the $[(B-V)_0,\,M_V]$-diagram, including the RGB, the failure of
the models to match the observed giants in the left-hand panel of
Fig.~\ref{fig:fig15} cannot be due to problems with just the color
transformations.  (Whatever is causing the mismatch between theory and the
$B-V$ colors may also be occurring in 47 Tuc, given that similar discrepancies
are evident for this system --- see Fig.~\ref{fig:fig14}).

Indeed, one cannot say very much about the reliability of different color
transformations using consistency arguments (i.e., from how well isochrones 
reproduce observations on different color planes) when GCs of apparently quite
similar metallicities and ages, like NGC$\,$1851 and M$\,$5, can have
significant variations of \delcol, which is the difference in color between the
turnoff and the RGB when measured at a fixed value of $\delta\,V$ brighter than
the turnoff.  Because M$\,$5 has a smaller value of \delcol\ than NGC$\,$1851,
it will not be possible to obtain satisfactory fits of isochrones to the RGBs
of both clusters on different CMDs.  It is surprising, however, that the models
appear to reproduce the RGB of NGC$\,$1851 so well on the
$[(V-I)_0,\,M_V]$-plane while providing a comparably good fit to the giant
branch of M$\,$5 on the $(B-V)_0,\,M_V]$-plane (and vice versa for the cluster
MS fiducials).  Although Figs.~\ref{fig:fig15} and~\ref{fig:fig16} provide good
support for the various color--$\teff$ relations applicable to dwarf stars
having [Fe/H] $\approx -1.4$, the color transformations that are applicable to
giants of the same metallicity cannot be assessed until we have a good
understanding of the cause(s) of the difference in \delcol\ between NGC$\,$1851
and M$\,$5.  

\subsection{M$\,$3 ([Fe/H] $\approx -1.6$)}
\label{subsec:m3}

M$\,$3 is often considered to be the prototype of the intermediate-metal-poor
group of GCs because it has a normal HB for its metallicity (relative to the
expectations from stellar models) and its giants do not show the same degree
of chemical abundance anomalies that are seen in many other systems of similar
metallicity (e.g., giants in M$\,$13, but not in M$\,$3, have super-low oxygen
abundances; see \citealt{ksl92}).  However, estimates of its iron content have
varied considerably over the years, from [Fe/H] $= -1.66$ (\citealt{zw84}) to
$-1.34$ (\citealt{cg97}) to $-1.50$ (\citealt{ki03}; \citealt{cbg09}).  As in
the case of NGC$\,$1851, comparisons of isochrones with the CMD of M$\,$3 tend
to favor a value near the low end of this range (e.g., \citealt{van00}).  This
has not changed, despite on-going improvements to both the photometric data and
the theoretical models.

In Figure~\ref{fig:fig17}, a VR2010 isochrone for [Fe/H] $=-1.60$ is compared
with $BVI$ observations for M$\,$3, on the assumption of $E(B-V) = 0.013$
(Schlegel et al.~1998) and $(m-M)_V = 15.00$, which is derived from a 
main-sequence fit of the cluster photometry to the isochrones on the 
$[(V-I)_0,\,M_V]$-diagram, and which agrees well with recently published
estimates (see, e.g., \citealt{rcp99}, \citealt{ccc05}).  Our $B$ photometry is
not as deep as those for $V$ and $I$; consequently, the cluster's principal
sequence is not as well defined in the left-hand panel as in the right-hand
panel.  Still, it is a little disconcerting that the CRMBA-transformed isochrone
is $\sim 0.02$ mag redder in the left-hand panel than it should be to provide a
completely consistent interpretation of both the $BV$ and $VI$ observations.
According to the comparison of our isochrones with the most metal-deficient
subdwarfs in Fig.~\ref{fig:fig13} (the filled circles), the $B-V$ colors of
the models appear to be, if anything, slightly too blue (not too red).  Whether
this discrepancy is due, in part, to the assumption of incorrect metal
abundances for M$\,$3 or to calibration errors is not known.

The fact that the isochrones provide such a good match to the entire RGB of
M$\,$3 on both color planes reinforces our contention that the difficulties
that were encountered when analyzing observations of NGC$\,$1851 and M$\,$5 are
not due simply to problems with the color--$\teff$ relations, but instead
suggest that the models themselves are lacking in some fundamental way (see the
discussion in the previous section).  In fact, NGC$\,$1851 has recently been
discovered to have a double SGB (\citealt{mbg08}), which is independently
confirmed by our data (see \citealt{ste09}; \citealt{msp09}).  The second
(fainter) SGB is not evident in our plots because, to maximize the clarity of
the comparison between the observations and the theoretical loci, we have 
plotted only a small, representative sample of photometric measurements
selected to have the highest accuracy and precision.  The relatively sparse
population on the second SGB is not numerous enough to be evident in this
sample.  It has also been discovered recently that NGC$\,$1851 shows chemical
abundance anomalies, which are consistent with the hypothesis that the present
stellar populations in this GC formed out of gas ejected by the
asymptotic-giant-branch stars from a previous generation (\citealt{ygd09}).
Thus, there is some justification for believing that the assumed mix of heavy
elements in the isochrones which we have fitted to the CMDs of (at least)
NGC$\,$1851 is not realistic.

\subsection{M$\,$92 ([Fe/H] $\approx -2.4$)}
\label{subsec:m92}

M$\,$92 presents us with a somewhat different puzzle.  As shown in 
Figure~\ref{fig:fig18}, isochrones for [Fe/H] $= -2.40$, with the usual
enhancement in the abundances of the $\alpha$-elements, are able to reproduce
the detailed CMD morophology of this cluster on the $B-V$ and $V-R$ color
planes {\it reasonably} well if canonical estimates of the reddening, $E(B-V)
= 0.023$ (Schlegel et al.~1998), and distance modulus, $(m-M)_V = 14.62$, are
assumed.  (The latter is obtained by matching the SGB of M$\,$92 to the field
subgiant, HD$\,$140283, which has close to the same metallicity as the GC;
see \citealt{vrm02}.)  However, to obtain a fully consistent fit of the
isochrones to the cluster observations on the $[(V-I)_0,\,M_V]$-plane as well,
a significant blueward correction to the isochrone $V-I$ color indices is
required (as indicated).  Of all the star clusters considered in this
investigation, this is the only one where the models fail to match the observed
$V-I$ colors for MS stars, on the assumption of what we consider to be best
estimates of the metallicity, reddening, and distance, without having to
correct the isochrones in some way.

It is worth noting that the adopted metallicity agrees well with recent
determinations of the [Fe/H] value of M$\,$92 from high-resolution spectroscopy,
which lie in the range of $-2.35$ to $-2.40$ (\citealt{ki03}; \citealt{cbg09}).
Had models for [Fe/H] $\approx -2.2$ been selected, to be in better agreement
with the original \citet{cg97} and \citet{zw84} estimates of $-2.16$ and
$-2.24$, respectively, they would have been offset to the red by even larger
amounts.  What makes M$\,$92 especially intriguing is that (i) excellent
consistency on all three color planes is obtained if a small zero-point 
correction is applied to the model $V-R$ colors (an amount that is well within
calibration uncertainties) and a constant offset of $\approx -0.025$ mag is
applied to the $V-I$ colors, and (ii) even larger adjustments (in the same
direction) appear to be necessary to match the RGB segments of the same
isochrones to the M$\,$92 giants on the $[(V-J)_0,\,M_V]$- and
$[(V-K)_0,\,M_V]$-diagrams (\citealt{bsv10}).  In the case of the near-IR CMDs,
the MS stars can be fitted quite well by the isochrones, but the giants are as
much as $\sim 0.12$ mag bluer in $V-K$ than the isochrones.

Careful inspection of Fig.~\ref{fig:fig18} reveals another anomaly; namely, that
the MARCS and CRMBA transformations for [Fe/H] $= -2.4$ yield fairly similar
$B-V$ colors but different $V-I$ colors (by $\sim 0.02$ mag) along the MS, which
is opposite to what was found at higher metallicities and opposite to what was
inferred from the subdwarfs.  (However, there are no subdwarfs in our sample
with [Fe/H] $\approx -2.4$, and only four stars with [Fe/H] $< -1.5$.  The most
metal-poor one is HD$\,$19445, which has [Fe/H] $\approx -2.0$; consequently,
these stars do not provide any constraint on the color--$\teff$ relations for
stars as metal-deficient as those in M$\,$92.)  Figure~\ref{fig:fig19} 
compares the isochrones for [Fe/H] $= -2.40$ and $-1.40$ that have been used
in this study.  At [Fe/H] $=-1.40$, the CRMBA- and MARCS-transformed 
isochrones predict nearly the same $V-I$ and $V-R$ colors, but the $B-V$
indices differ by $\sim 0.02$--0.03 mag at the same absolute magnitude.  At
[Fe/H] $= -2.40$, there are only small differences between the predicted $B-V$
colors, whereas the $V-R$ and $V-I$ colors for dwarf stars are, in turn,
$\approx 0.01$ and $\approx 0.02$ mag bluer, at a given $M_V$, than those
obtained from the MARCS color--$\teff$ relations.  Given the many uncertainties
at play, it is difficult to determine whether these findings are trustworthy.

The bottom line is that we are unable to obtain a fully consistent explanation
of the optical photometry of M$\,$92 (or the near-IR data, judging from the
work of \citealt{bsv10}).  Isochrones for the current best estimate of the
cluster metallicity are able to reproduce the observed $[(B-V)_0,\,M_V]$-diagram
quite well, and aside from a zero-point offset of $\approx 0.025$ mag, they
provide a good match to the entire $[(V-I),\,M_V]$-diagram.  However, the same
models apparently suffer from systematic errors when compared with $V-J,\,V$
and $V-K_S,\,V$ observations (see Brasseur et al.), insofar as they provide a
reasonable fit to the MS stars, but not to the giants.  There is no way in which
the evolutionary calculations could be modified to reconcile these conflicting
indications.  Consequently, errors from several sources --- the models, the
color--$\teff$ relations, the photometric data themselves, and perhaps the
basic cluster parameters (reddening, distance, and chemical composition) ---
must be conspiring to cause the problems described above.  A resolution of
the M$\,$92 conundrum must be left to future work.



\section{Summary}
\label{sec:conclude}

This investigation has examined how well up-to-date theoretical isochrones
that take the diffusion of helium into account are able to satisfy various
observational constraints when they are transformed from the theoretical to the
$(B-V,\,V)$-, $(V-R,\,V)$-, and $(V-I,\,V)$-diagrams using the CRMBA, MARCS, and
VC03 color transformations.  In fact, the differences between the three sets
of color--$\teff$ relations that have been considered are relatively minor,
especially in the case of stars having close to the solar metallicity.  Our
consideration of the Hyades and M$\,$67 has shown that, aside from the need
to apply a blueward correction of $\sim 0.02$ mag to the $V-I$ colors given
by VC03, isochrones are able to reproduce the observed CMDs very well,
independently of which of the three color transformations are used.  Only at
$M_V \gta 6.5$ (or fainter in the case of $V-I$) do the models fail to match
the observed fiducial sequences: the MARCS-tranformed isochrones deviate to
the blue, possibly because of insufficient blanketing in the model atmospheres
for cool stars on which they are based, while the opposite is found in the
case of the CRMBA-transformed isochrones, which is likely due to the limitations
of the analytic expressions used to present these empirical transformations.
However, we were unable to obtain perfectly consistent fits of isochrones to
$BV$ and $VI$ photometry for NGC$\,$6791.  Whether this is indicative of a
problem with the color--$\teff$ relations for super-metal-rich stars, the
assumed chemical abundances, or any of the other factors that play a role in
such comparisons is not known.

One of the most striking results of this work is that the hot $\teff$ scale
derived by CRMBA is in remarkable agreement with that predicted by stellar
models.  Comparisons of isochrones with 33 nearby subdwarfs having [Fe/H]
values between $-2.0$ and $-0.5$, with well-determined $M_V$ values from
{\it Hipparcos}, have shown that the mean metallicites and temperatures that
are inferred for the stars from their locations relative to the models on the
$(\log\teff,\,M_V)$-plane agree with those given by CRMBA to within
$\delta$[Fe/H] $=0.05$ dex, and $\delta\teff = 10$~K, respectively (see
Fig.~\ref{fig:fig10}), which is obviously well within the uncertainties.
Not surprisingly, because the CRMBA color--$\teff$
relations are based on a large sample of stars that includes the 33 subdwarfs,
similar consistency is found on the $B-V$, $V-R$, and $V-I$ color planes.  When
the same comparisons are made using isochrones that employ the MARCS 
transformations, the predicted $B-V$ colors are found to be too blue by about
0.03 mag, while the inferred $V-R$ and $V-I$ colors agree quite well
with those observed.  Why the MARCS $(B-V)$--$\teff$ relations would be more
problematic for metal-deficient stars than for those having [Fe/H\ $\gta 0.0$
is not clear, but if the MARCS transformations to $V-R$ and $V-I$ {\it are}
more trustworthy, then the $\teff$ scale implied by the MARCS model atmospheres
is not significantly different from the empirical one derived by CRMBA.  Both
give warmer temperatures by $\sim 75$--120~K than, e.g., Alonso et al.~(1996)
and \citet{cps07}.

\begin{figure}
\plotone{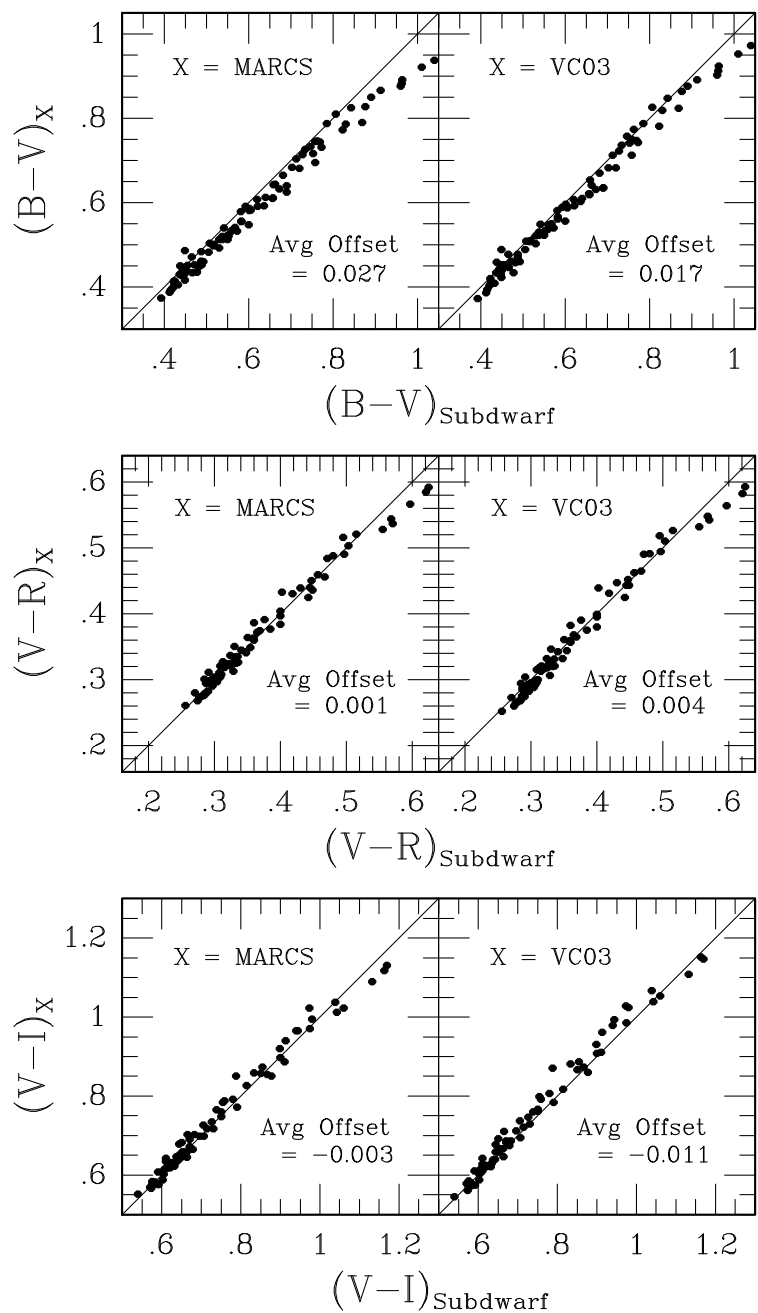}
\caption{Comparison of the observed subdwarf colors with those predicted by
MARCS and VC03 color transformations on the assumption of the subdwarf
properties (i.e., $\teff$, $\log\,g$, and [Fe/H]) determined by CRMBA.  The
noted offsets are in the sense ``observed minus predicted".}
\label{fig:fig8}
\end{figure}

\begin{figure}
\plotone{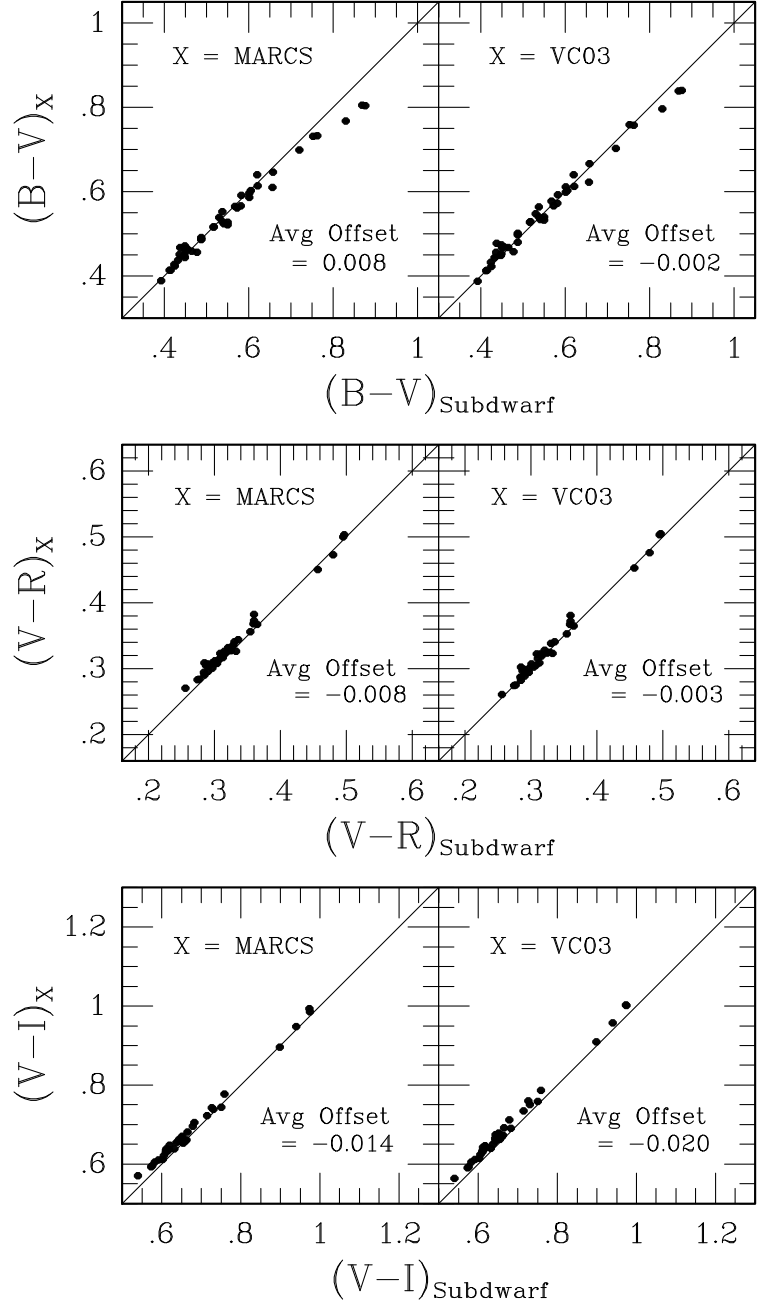}
\caption{As in the previous figure, except that the $\teff$, $\log g$, and
[Fe/H] values derived by Gratton et al.~are assumed in calculating the
corresponding MARCS and VC03 colors.}
\label{fig:fig9}
\end{figure}

\begin{figure}
\plotone{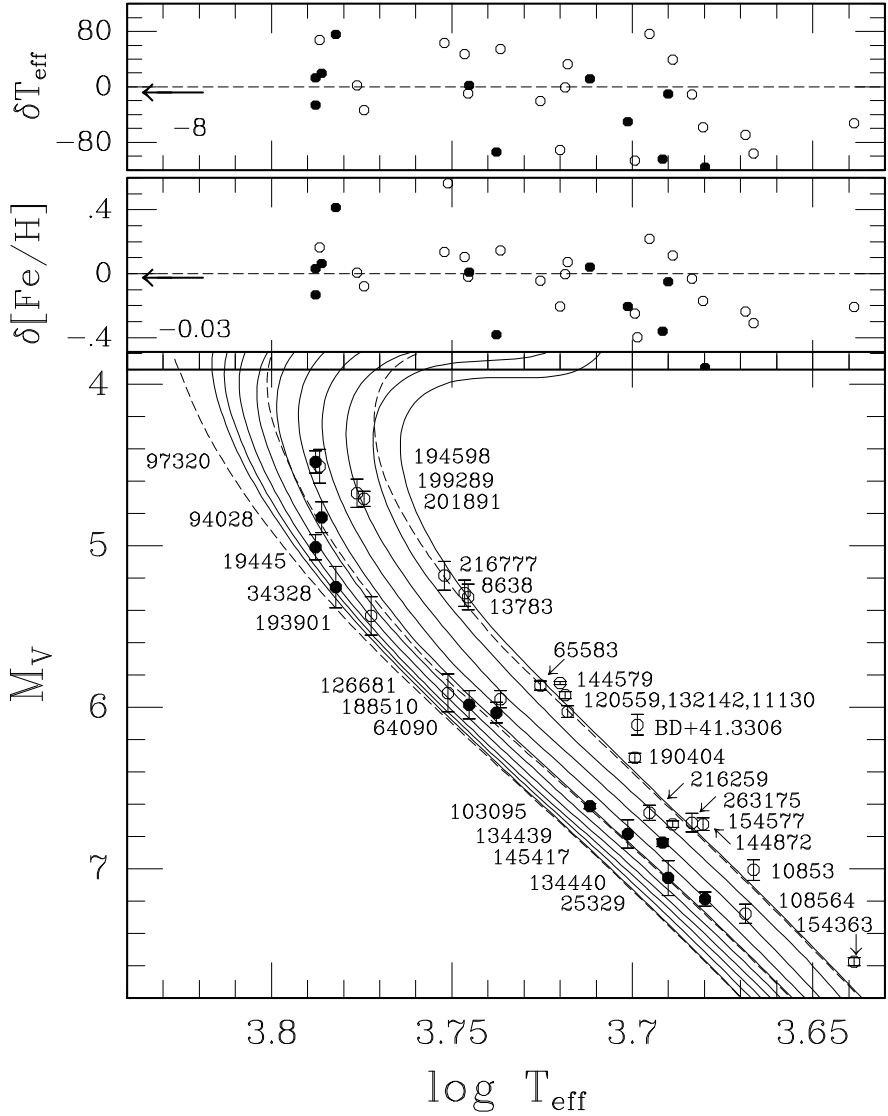}
\caption{{\it Lower panel:}~Superposition of the properties of local subdwarfs
having well determined $M_V$ values onto a set of 12 Gyr isochrones for [Fe/H]
values from $-2.40$ to $-0.60$, in 0.2 dex steps (from left to right).
Isochrones for $-2.4$, $-1.4$, and $-0.60$, but for an age of 10 Gyr, are
plotted as dashed curves.  Filled and open circles represent subdwarfs having
[Fe/H] $\le -1.2$ and $> -1.2$ (according to CRMBA), respectively.  The error
bars on the points depict the $1 \sigma$ uncertainties in the $M_V$ values as
derived from {\it Hipparcos} parallaxes (\citealt{vl07}).  The stars are
identified by their ``HD" numbers, except for BD$+41\,3306$.  {\it Middle panel:}
Plotted as a function of $\log\teff$, the difference between the CRMBA 
estimate of [Fe/H] for each star and that inferred from the interpolated (or
extrapolated) isochrone that matches its location on the 
$(\log\teff,\,M_V)$-diagram in the lower panel.  {\it Upper panel:} The
difference in $\teff$ that would need to be applied to each subdwarf in order
to achieve perfect consistency of its position in the lower panel.  The numbers
and arrows in the middle and upper panels give the mean values of $\delta$[Fe/H]
and $\delta\teff$, respectively, that were computed using all stars.}
\label{fig:fig10}
\end{figure}

\begin{figure}
\plotone{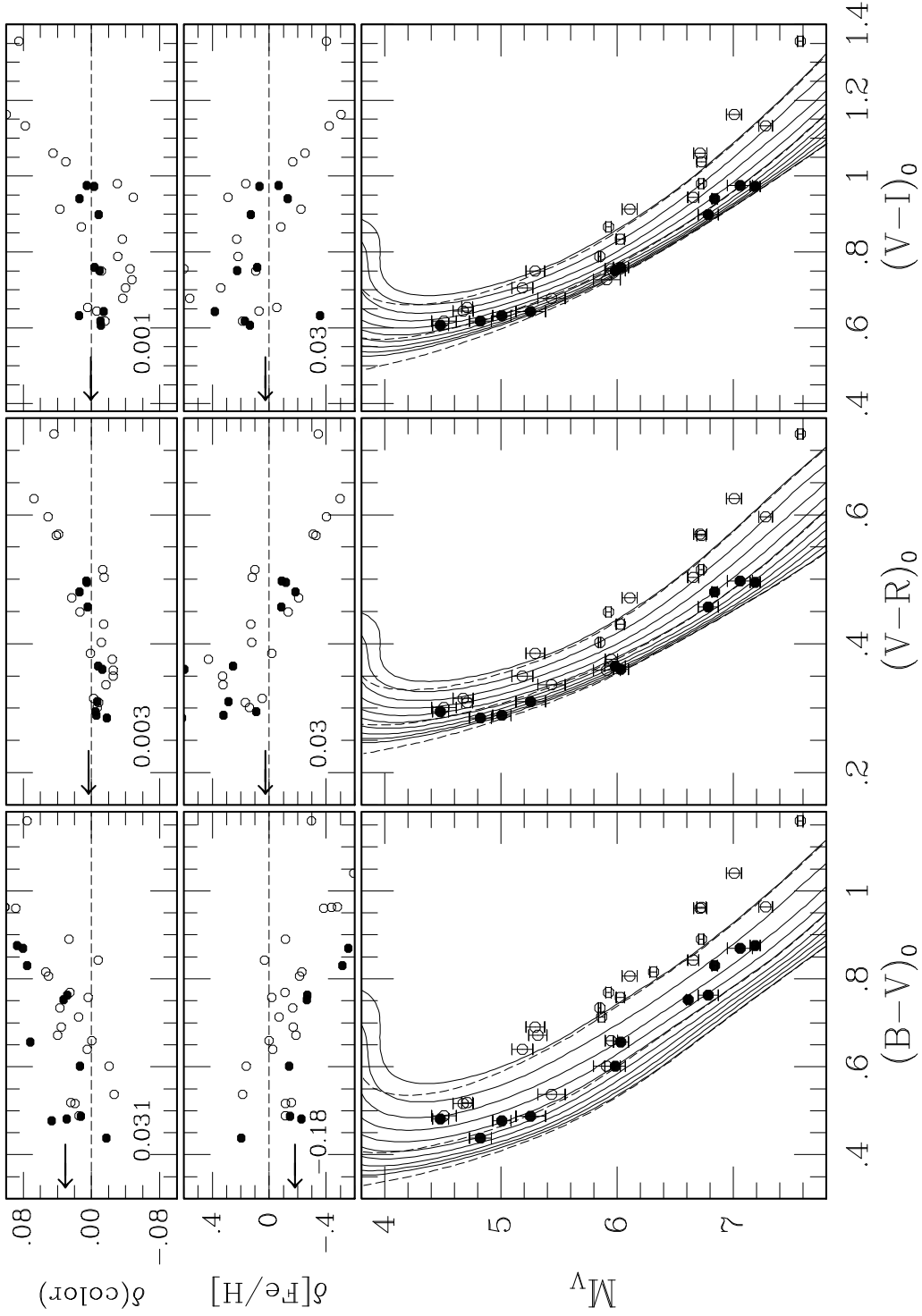}
\caption{Similar to the previous figure, except that the subdwarfs are compared
with the isochrones that have been transposed to the $B-V$, $V-R$, and $V-I$
color planes using the MARCS color--$\teff$ relations.  The subdwarf [Fe/H]
values and colors are taken from the study by CRMBA.}
\label{fig:fig11}
\end{figure}

\begin{figure}
\plotone{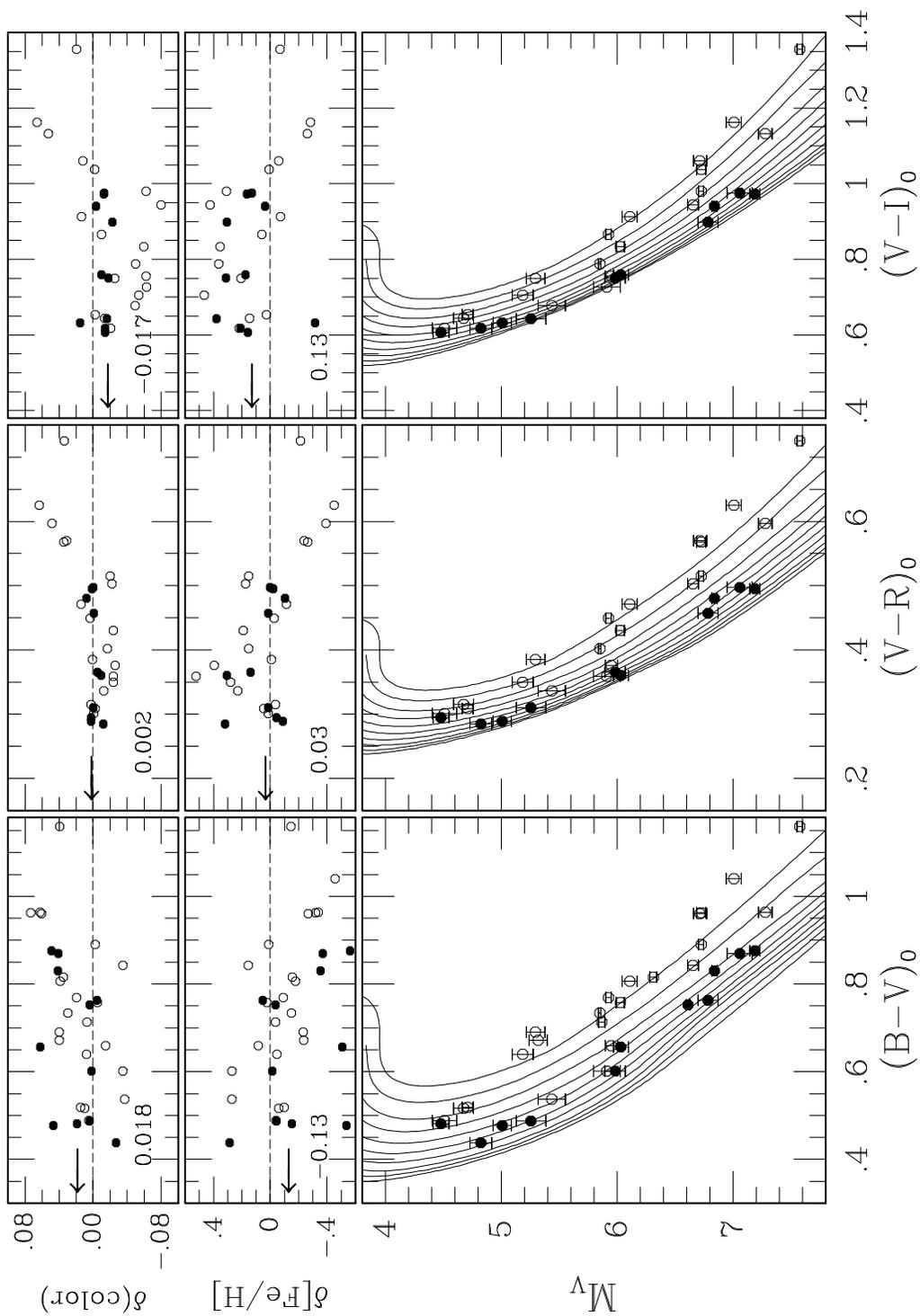}
\caption{As in the previous figure, except that the VC03 color--$\teff$
relations have been used to transpose the isochrones to the various observed
planes and the dashed curves have been omitted.}
\label{fig:fig12}
\end{figure}

\begin{figure}
\plotone{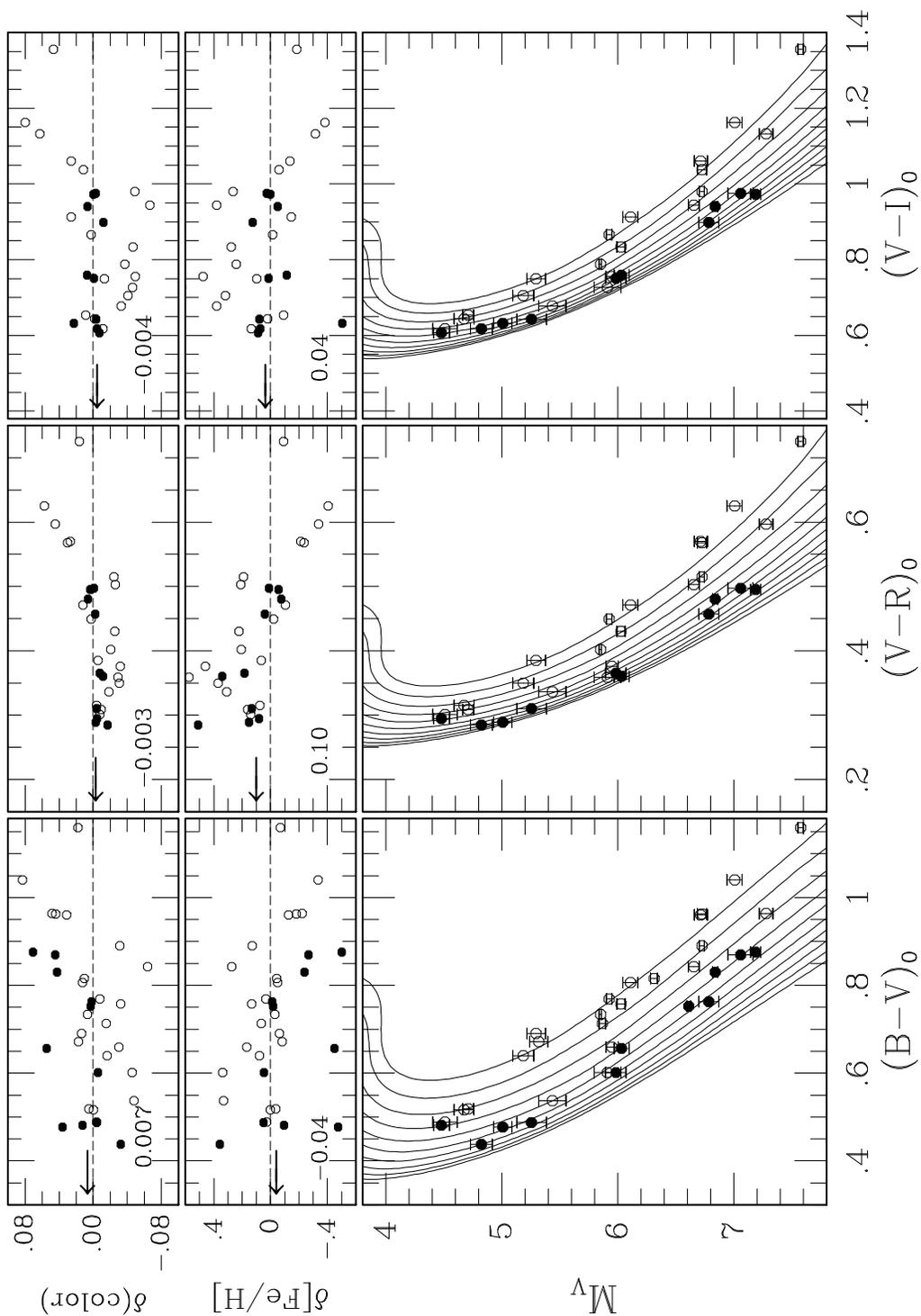}
\caption{As in the previous figure, except that the empirical color--$\teff$
relations given by CRMBA for dwarf and subgiant stars have been used to
transpose the isochrones to the various observed planes.}
\label{fig:fig13}
\end{figure}

While these results depend quite critically on the adopted [Fe/H] values of the
subdwarfs (something that should be kept in mind), color--$\teff$ relations 
are not, by themselves, very dependent on the metal abundance (especially at
lower metallicities).  Consequently, it is reassuring to find that similar
conclusions are reached regarding the MARCS transformations when the colors of
$\sim 100$ local subdwarfs and subgiants are compared with those obtained by
interpolating in the MARCS color tables for the values of $\teff$, $\log\,g$,
and [Fe/H] given by CRMBA for those stars.  As shown in Fig.~\ref{fig:fig8},
the predicted $B-V$ indices are $\approx 0.03$ mag too blue, in the mean,
while the predicted $V-R$ and $V-I$ colors agree very well with those observed.
When subjected to the same tests, the VC03 transformations fare comparably, or
less, well (depending on the color index considered), though they do enable
models to provide satisfactory matches to the lower-MS slopes of observed CMDs.
However, except for providing some guidance concerning the variation of the
colors of cool MS stars with temperature, the VC03 color--$\teff$ relations are
no longer very useful: they have effectively been superseded by the new MARCS
transformations.  Because they provide very good consistency on {\it all} color
planes, the empirical transformations of CRMBA are the preferred ones to use
for dwarf and SGB, but not necessarily lower RGB, stars.

\begin{figure}
\plotone{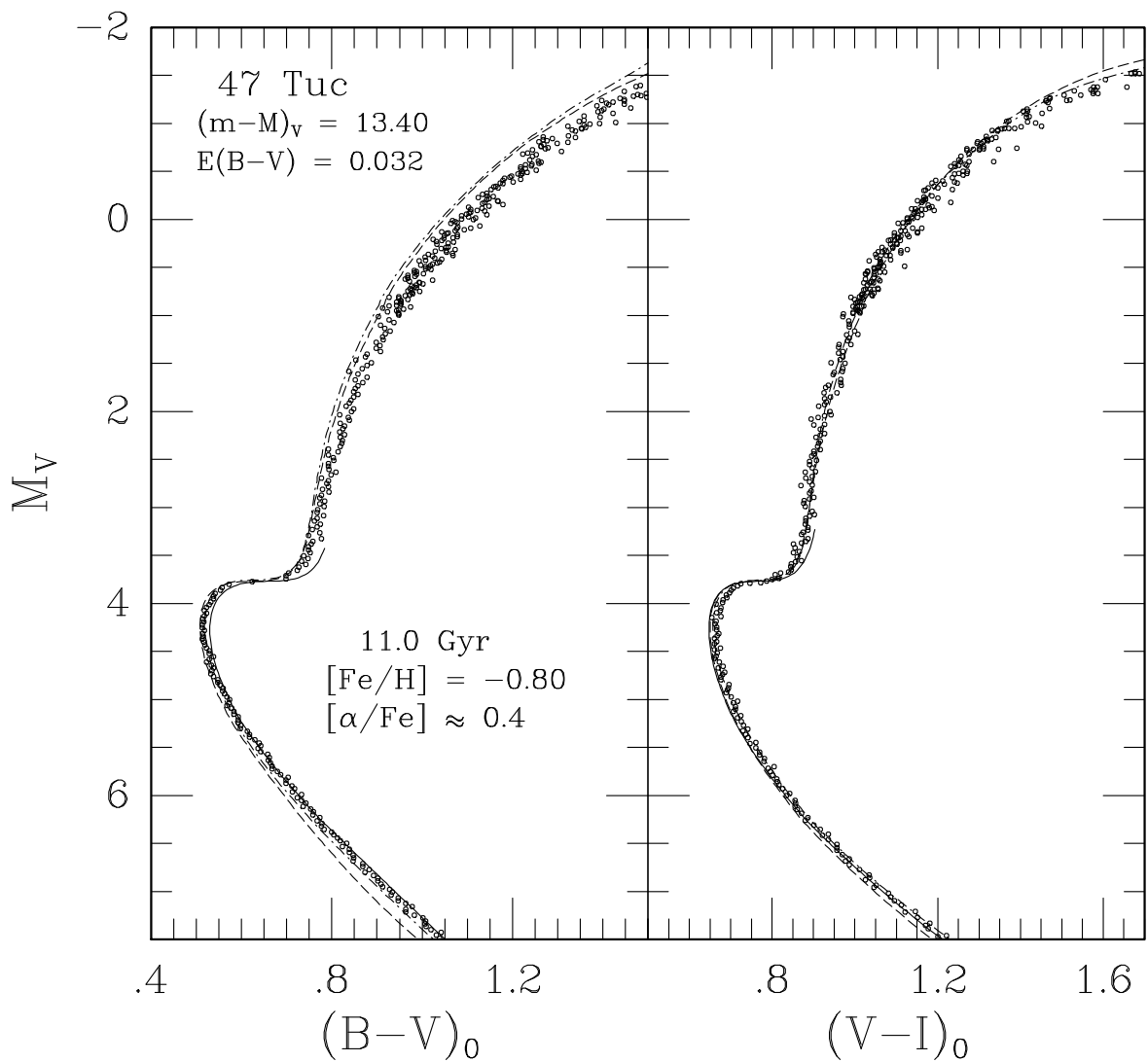}
\caption{Comparison of an 11.0 Gyr isochrone for [Fe/H] $= -0.8$, [$\alpha$/Fe]
$\approx 0.4$, and $Y = 0.25$ with photometry for 47 Tucanae from \citet{bs09}.
The solid, dashed, and dot-dashed curves assume, in turn, the CRMBA, MARCS, and
VC03 color transformations.  The predicted and observed turnoff colors agree to
within 0.02 mag, with the solid curves showing the largest blueward and redward
offsets in both the right- and left-hand panels, respectively.}
\label{fig:fig14}
\end{figure}

\begin{figure}
\plotone{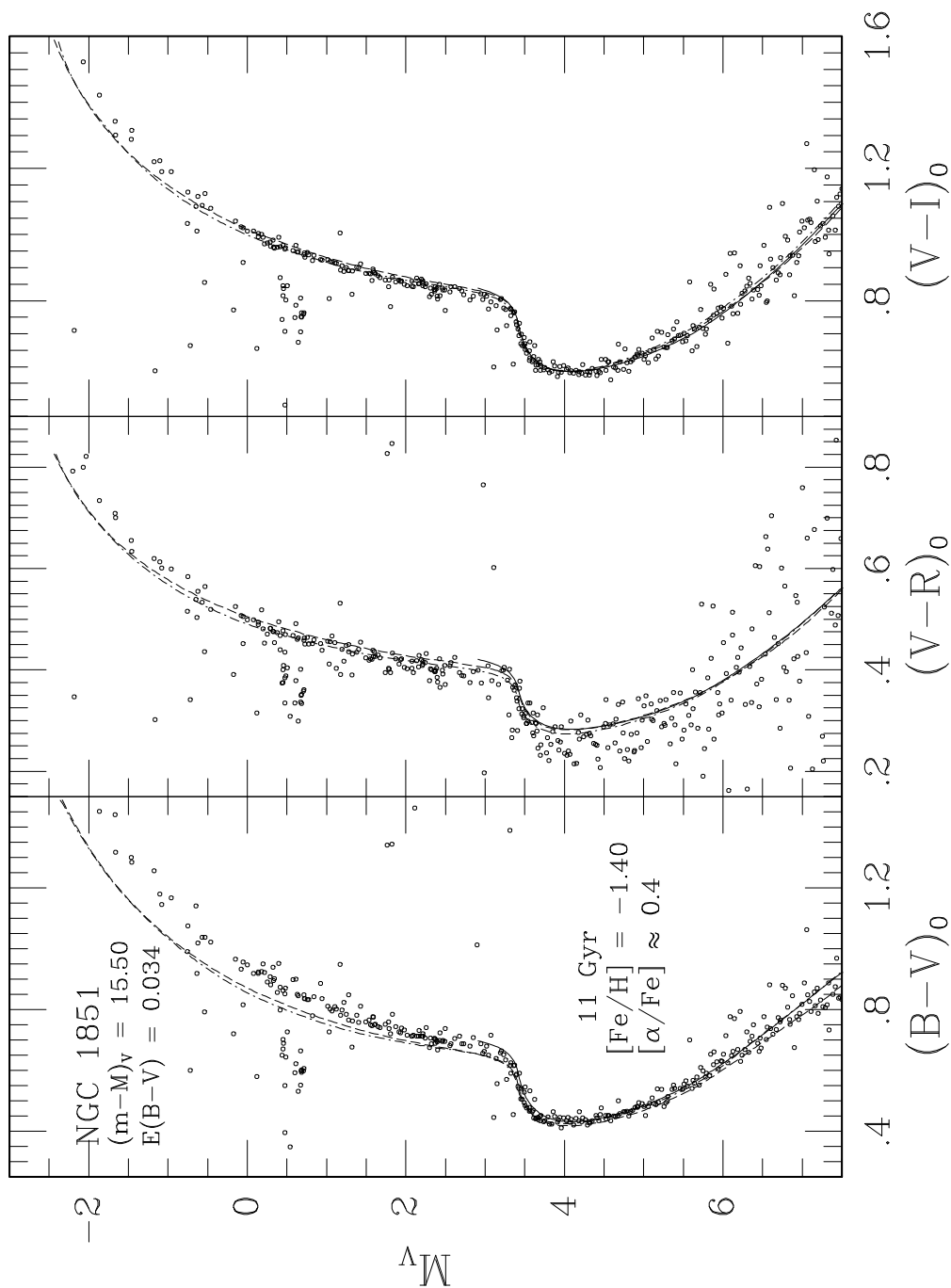}
\caption{Comparison of an 11.0 Gyr isochrone for [Fe/H] $= -1.40$,
[$\alpha$/Fe] $\approx 0.4$, and $Y = 0.25$ with the latest calibration of
photometry for NGC$\,$1851 from the \citet{ste00} database.  The solid, dashed,
and dot-dashed curves assume, in turn, the CRMBA, MARCS, and VC03 color
transformations.  In general, the predicted and observed turnoff colors agree
to within 0.015 mag.}
\label{fig:fig15}
\end{figure}

\begin{figure}
\plotone{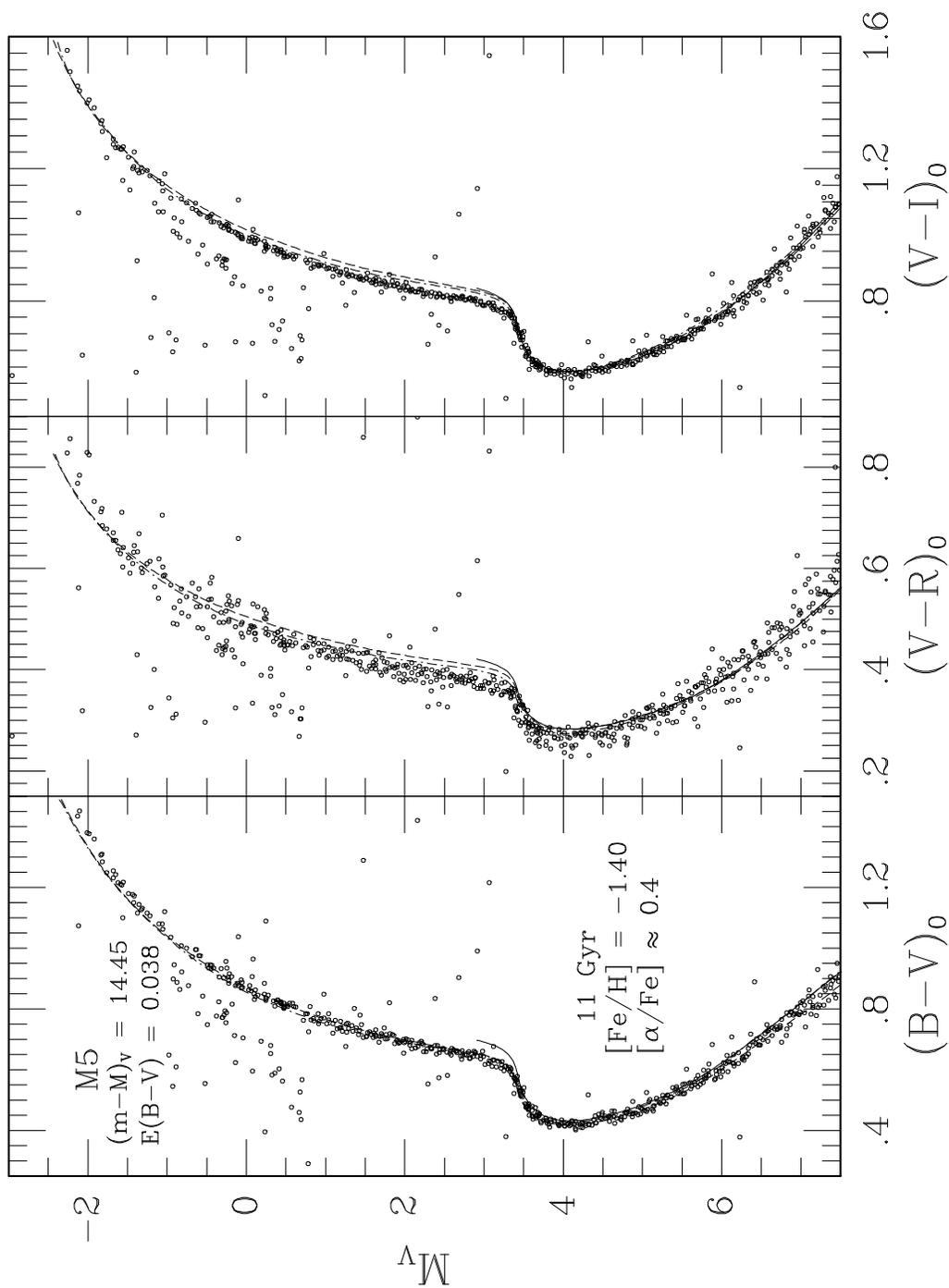}
\caption{As in the previous figure, except that the same isochrones are 
compared with our photometry for M$\,$5.  Here as well, the predicted and
observed turnoff colors agree to within 0.015 mag.}
\label{fig:fig16}
\end{figure}

\begin{figure}
\plotone{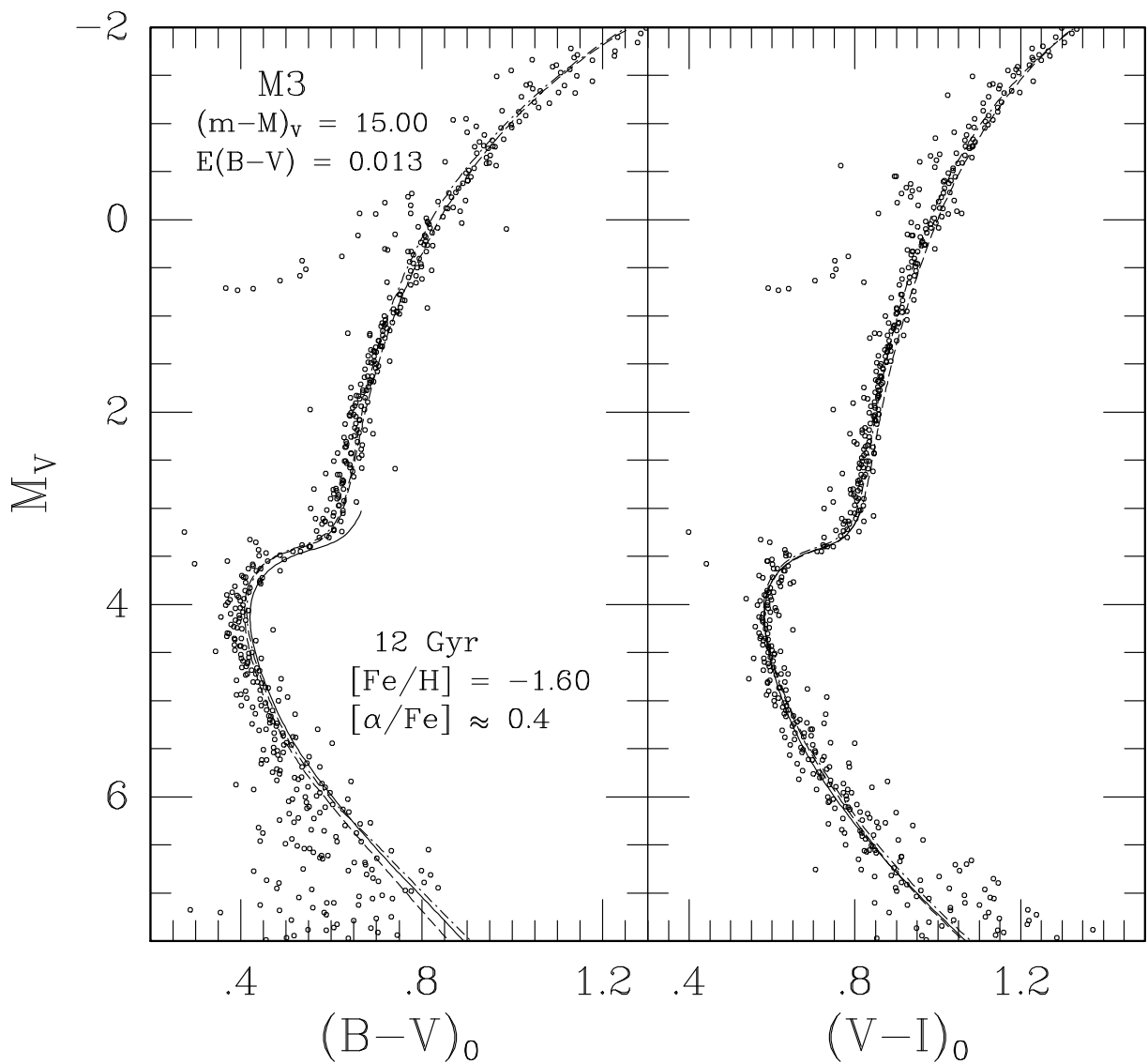}
\caption{As in the previous figure, except that a 12.0 Gyr isochrone for
a slightly lower metallicity (as noted) is compared with observations of 
M$\,$3.  Whereas there is excellent consistency of the predicted and turnoff
colors in the right-hand panel, the observed turnoff is 0.02--0.025 mag
bluer than that predicted by the solid curve, with somewhat smaller offsets
in the case of the other isochrones, though in the same sense.}
\label{fig:fig17}
\end{figure}

\begin{figure}
\plotone{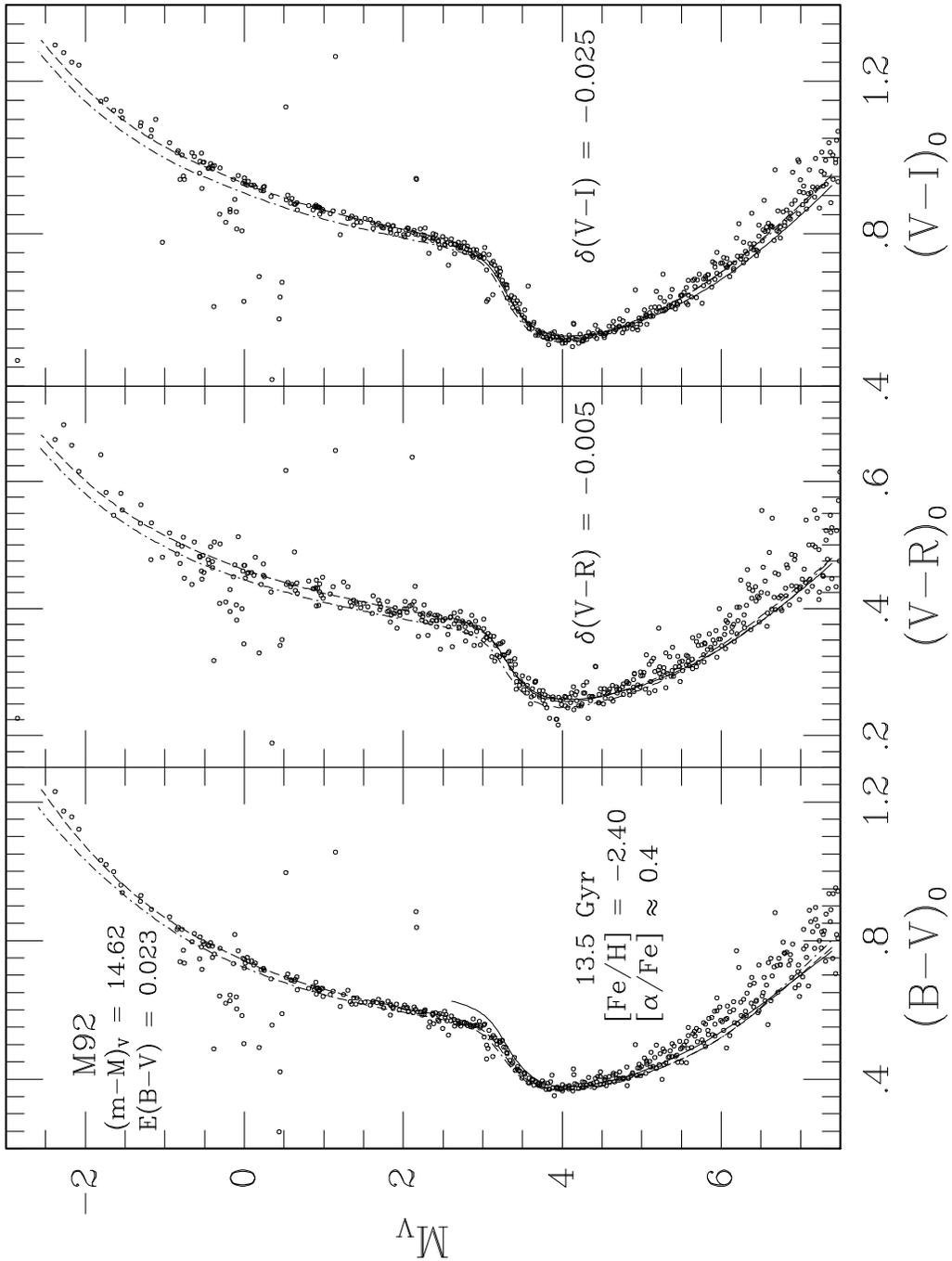}
\caption{As in the previous figure, except that a 13.5 Gyr isochrone for
[Fe/H] $= -2.40$ is compared with observations of M$\,$92.  Aside from the
large zero-point offset that has been applied to the observations in the
right-hand panel, as indicated, there is good consistency of the predicted and
observed turnoff colors (i.e., to within 0.015 mag).}
\label{fig:fig18}
\end{figure}

\begin{figure}
\plotone{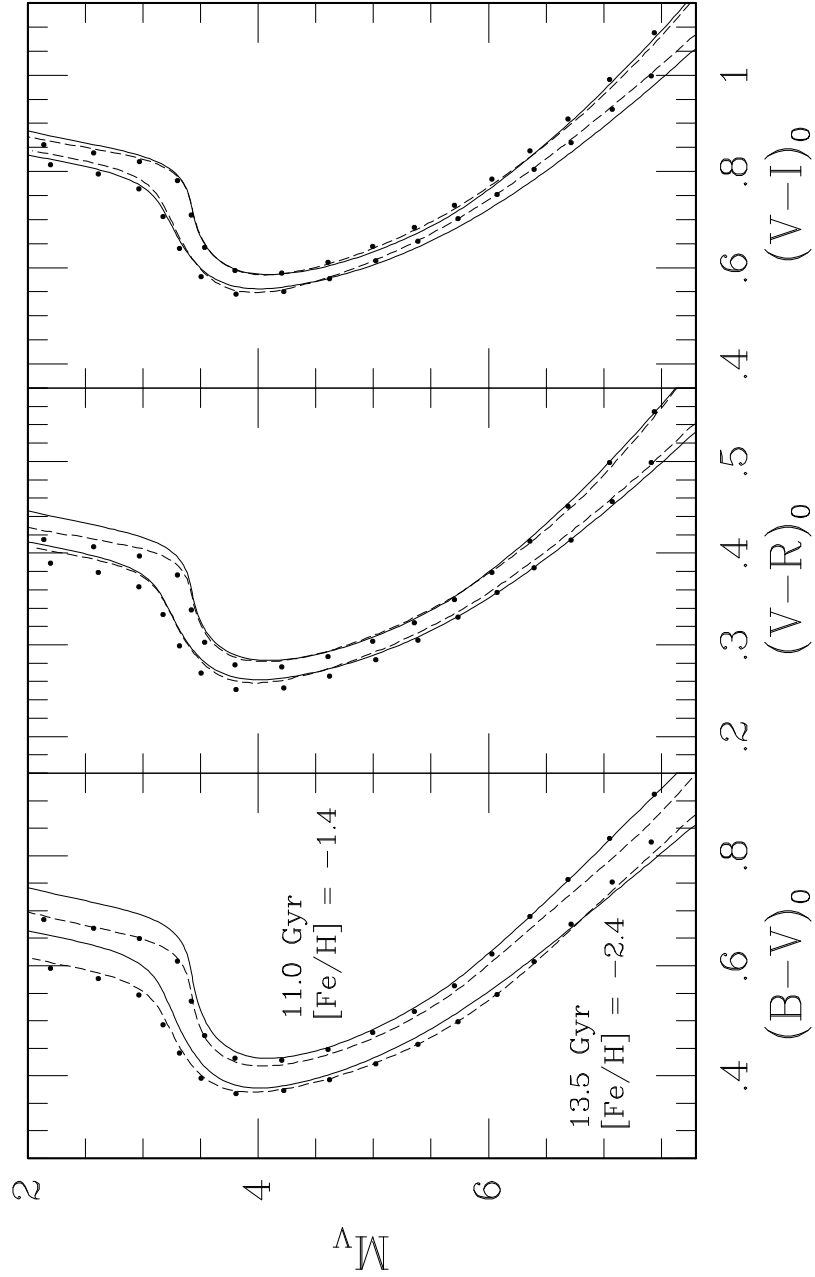}
\caption{Comparisons of isochrones for the indicated ages and [Fe/H] values
using the CRMBA (solid curve), MARCS (dashed curve), and VC03 (dotted curve)
color transformations.}
\label{fig:fig19}
\end{figure}

Although we are able to obtain reasonably consistent fits of the same isochrones
to the dwarf and SGB populations of GCs on different color-magnitude planes ---
when well constrained estimates of reddening, distance, and metallicity are
assumed --- the cluster RGBs are much more problematic.  In the case of M$\,$3
(see Fig.~\ref{fig:fig17}), the models reproduce the observed $(B-V)_0$ and
$(V-I)_0$ colors of the cluster giants quite well, but more often than not,
isochrones are able to reproduce the $VI$ photometry, but not the $BV$
observations, along the cluster giant branch (e.g., see Fig.~\ref{fig:fig14}
for 47 Tucanae and Fig~\ref{fig:fig15} for NGC$\,$1851), {\it or} vice versa
(e.g., see Figs.~\ref{fig:fig16} and~\ref{fig:fig18} regarding M$\,$5 and
M$\,$92, respectively).  In view of recent work which has shown that the
location of the RGB on the H-R diagram is a sensitive function of the mix of
heavy elements (\citealt{dcf07}; VR2010), we suspect that differences between
the assumed and actual metal abundances may be the main cause of the noted
difficulties.  Indeed, the temperatures of MS stars can also be affected by
variations in the abundances of such elements as Mg and Si, though to a lesser
extent, which clearly complicates the interpretation of GC CMDs.  Follow-up
studies must be undertaken to explore how color--$\teff$ relations are modified
by such chemical abundance variations and to determine whether or not the
resultant transformations lead to improved fits of theoretical isochrones to
observations of GCs (on all color planes) compared with those presented here.

We note, finally, that it appears to be impossible to reconcile stellar models
with all of the available photometric data for M$\,$92.  The $BV(RI)_C$
observations alone do not pose a serious problem, as isochrones for $Y=0.25$,
[Fe/H] $= -2.40$, and [$\alpha$/Fe] $\approx 0.4$ match the entire CMD rather
well on the $B-V$, $V-R$, and $V-I$ color planes, provided that the predicted
$V-I$ colors are adjusted to the blue by a small, constant amount (0.025 mag).
While it is odd that the $V-I$ indices appear to be more problematic than the
$B-V$ colors, what is really unexpected is that the same isochrones show large,
and systematic, discrepancies when compared with $V-J$ and $V-K_S$ observations
of M$\,$92 --- which are from the 2MASS catalogue in the case of the cluster
giants (see \citealt{bsv10}).  The isochrones fit the MS and the upper RGB
satisfactorily, but they are too red by $\delta(V-K_S) \approx 0.12$ mag just
above the base of the giant branch.  Is it possible that the MARCS model
atmospheres for very metal-deficient upper MS and RGB stars are too bright in
the near-IR?  It seems unlikely, but this and other possible explanations need
to be investigated.  As a footnote to the main results of this study, it is
worth pointing out that we find a significant dependence of GC ages on
metallicity.  Isochrones that faithfully reproduce the properties of local
subdwarfs with accurate distances from {\it Hipparcos} predict that the ages
of these systems vary from $\approx 13.5$ Gyr at [Fe/H] $\approx -2.4$ (M$\,$92)
to $\approx 11$ Gyr at [Fe/H] $\approx -0.8$ (47 Tuc).  A similar variation was
found by \cite{van00}, who used empirically constrained HB luminosities to
establish the GC distance scale.

\acknowledgements
This work has been supported by the Natural Sciences and Engineering
Research Council of Canada through a Discovery Grant to DAV.

\clearpage

\clearpage

\clearpage

\clearpage

\clearpage

\clearpage

\clearpage

\clearpage

\clearpage

\clearpage

\clearpage

\clearpage

\clearpage

\clearpage

\clearpage

\clearpage

\clearpage

\clearpage

\clearpage


\newpage
 
\begin{thebibliography}{}

\bibitem[Alonso, Arribas, \& Martinez-Roger(1996)]{aam96}
Alonso, A., Arribas, S., \& Martinez-Roger, C.~1996, A\&A, 117, 227

\bibitem[Alonso, Arribas, \& Martinez-Roger(1999)]{aam99}
Alonso, A., Arribas, S., \& Martinez-Roger, C.~1999, A\&AS, 140, 261

\bibitem[Asplund et al.(2005)]{ags05}
Asplund, M., Grevesse, N., Sauval, S.~J., Allende Prieto, C., \& Blomme, 
  R.~2005, A\&A, 431, 693

\bibitem[Bahcall et al.(2005)]{bbs05}
Bahcall, J.~N., Basu, S., Pinsonneault, M., \& Serenelli, A.~M.~2005, ApJ, 618,
1049

\bibitem[Barklem et al.(2002)]{bsa02}
Barklem, P.~S., Stempels, H.~C., Allende Prieto, C., Kochukhov, O.~P.,
Piskunov, N., \& O'Mara, B.~J.~2002, A\&A, 385, 951

\bibitem[Bergbusch \& Stetson(2009)]{bs09}
Bergbusch, P.~A., \& Stetson, P.~B.~2009, AJ, 138, 1455

\bibitem[Bergbusch \& VandenBerg(2001)]{bv01}
Bergbusch, P.~A., \& VandenBerg, D.~A.~2001, ApJ, 556, 322

\bibitem[Bessell(1990a)]{bes90a}
Bessell, M.~S.~1990a, A\&AS, 83, 357

\bibitem[Bessell(1990b)]{bes90b}
Bessell, M.~S.~1990b, PASP, 102, 1181

\bibitem[Bessell(1995)]{bes95}
Bessell, M.~S.~1995, PASP, 107, 672

\bibitem[Bessell(2005)]{bes05}
Bessell, M.~S.~2005, ARA\&A, 43, 293

\bibitem[Bessell, Castelli, \& Plez(1998)]{bcp98}
Bessell, M.~S., Castelli, F., \& Plez, B.~1998, A\&A, 333, 231

\bibitem[Boesgaard, Jensen, \& Deliyannis(2009)]{bjd09}
Boesgaard, A.~M., Jensen, E.~E.,~C., \& Deliyannis, C.~P.~2009, AJ, 137, 4949

\bibitem[Bohlin(2007)]{boh07}
Bohlin, R.~C.~2007, in The Future of Photometric, Spectrophotometric, and
Polarimetric Standardization, ed.~C.~Sterken, ASP, Conf.~Ser., 364, 315

\bibitem[Brasseur et al.(2010)]{bsv10}
Brasseur, C., Stetson, P.~B., VandenBerg, D.~A., Casagrande, L., Bono, G., \&
Dall'Ora, M.~2010, AJ, submitted


\bibitem[Cacciari, Corwin, \& Carney(2005)]{ccc05}
Cacciari, C., Corwin, T.~M., \& Carney, B.~W.~2005, AJ, 129, 267

\bibitem[Calamida et al.(2007)]{cbs07}
Calamida, A., Bono, G., Stetson, P.~B., et al.~2007, ApJ, 670, 400

\bibitem[Caldwell et al.(1993)]{cca93}
Caldwell, J.~A.~R., Cousins, A.~W.~J., Ahlers, C.~C., van Wamelan, P., \&
Maritz, E.~J.~1993, SAAO Circ., 15, 1

\bibitem[Carretta et al.(2009)]{cbg09}
Carretta, E., Bragaglia, A., Gratton, R.~G., D'Orazi, V., \& Lucatello, S.~2009,
A\&A, 508, 695

\bibitem[Carretta \& Gratton(1997)]{cg97}
Carretta, E., \& Gratton, R.~G.~1997, A\&AS, 121, 95

\bibitem[Carretta et al.(2000)]{cgc00}
Carretta, E., Gratton, R.~G., Clementini, G., \& Fusi Pecci, F.~2000, ApJ, 533,
215

\bibitem[Casagrande(2009)]{cas09}
Casagrande, L.~2009, Mem.~Soc.~Astron.~Ital., 80, 727

\bibitem[Casagrande, Flynn, \& Bessell(2008)]{cfb08}
Casagrande, L., Flynn, C., \& Bessell, M.~2008, MNRAS, 389, 585

\bibitem[Casagrande, Portinari, \& Flynn(2006)]{cpf06}
Casagrande, L., Portinari, L., \& Flynn, C.~2006, MNRAS, 373, 13

\bibitem[Casagrande et al.(2010; hereafter CRMBA)]{crm10}
Casagrande, L., Ram\'irez, I., Mel\'endez, J., Bessell, M., \& Asplund,
M.~2010, A\&A, 512, 54~~~~(CRMBA)

\bibitem[Cayrel de Strobel, Crifo, \& Lebreton(1997)]{ccl97}
Cayrel de Strobel, G., Crifo, F., \& Lebreton, Y.~in {\it Hipparcos--Venice
'97}, ESA SP-402 (Noordwijk: ESA), p. 687

\bibitem[Cenarro et al.(2007)]{cps07}
Cenarro, A.~J., Peletier, R.~F., S\'anchez-Bl\'azques, P., et al.~2007, MNRAS,
374, 664

\bibitem[Christensen-Dalsgaard(2009)]{cd09}
Christensen-Dalsgaard, J.~2009, astro-ph (arXiv: 0912.1405v1)

\bibitem[Clem et al.(2004)]{cvg04}
Clem, J.~L., VandenBerg, D.~A., Grundahl, F., \& Bell, R.~A.~2004, AJ, 127,
1227


\bibitem[Clementini et al.(1999)]{cge99}
Clementini, G., Gratton, R.~G., Carretta, E., \& Sneden, C.~1999, MNRAS, 302,
22

\bibitem[Collet, Asplund, \& Trompedach(2007)]{cat07}
Collet, R., Asplund, M., \& Trompedach, R.~2007, A\&A, 469, 687

\bibitem[de Bruijne, Hoogerwerf, \& de Zeeuw(2001)]{dhd01}
de Bruijne, J.~H.~J., Hoogerwerf, R., \& de Zeeuw, P.~T.~2001, A\&A, 367, 111

\bibitem[Cassisi et al.(2007)]{cpp07}
Cassisi, S., Potekhin, A., Pietrinferni, A., Catalan, M., \& Salaris, M.~2007,
ApJ, 661, 1094

\bibitem[Dotter et al.(2007)]{dcf07}
Dotter, A., Chaboyer, B., Ferguson, J.~W., Lee, H.-C., Worthey, G., 
Jevremovi\'c, D., \& Baron, E.~2007, ApJ, 666, 403

\bibitem[Dotter et al.(2008)]{dcj08}
Dotter, A., Chaboyer, B., Jevremovi\'c, D., Kostov, V., Baron, E., \& Ferguson,
J.~W.~2008, ApJS, 178, 89

\bibitem[Edvardsson(2008)]{edv08}
Edvardsson, B.~2008, Physica Scripta, Vol.~T133, 014011

\bibitem[Graham(1982)]{gra82}
Graham, J.~A.~1982, PASP, 94, 244

\bibitem[Gratton, Carretta, \& Castelli(1996)]{gcc96}
Gratton, R.~G., Carretta, E., \& Castelli,F.~1996, A\&A, 314, 191

\bibitem[Gratton, Sneden, \& Carretta(2004)]{gsc04}
Gratton, R., Sneden, C., \& Carretta, E.~2004, ARA\&A, 42, 385

\bibitem[Grevesse \& Sauval(1998)]{gs98}
Grevesse, N., \& Sauval, A.~J.~1998, Sp.~Sci.~Rev., 85, 161

\bibitem[Grundahl et al.(2008)]{gch08}
Grundahl, F., Clausen, J.~V., Hardis, S., \& Frandsen, S.~2008, A\&A, 492, 171

\bibitem[Gustafsson et al.(2008)]{gee08}
Gustafsson, B., Edvardsson, B., Eriksson, K., J{\o}rgensen, U.~G., Nordlund,
 {\AA}., \& Plez, B.~2008, A\&A, 486, 951

\bibitem[Holweger \& M\"uller(1974)]{hm74}
Holweger, H., \& M\"uller, E.~A.~1974, Sol.~Phys., 39, 19

\bibitem[Joner \& Taylor(1988)]{jt88}
Joner, M.~D., \& Taylor, B.~J.~1988, AJ, 96, 218

\bibitem[Kaluzny \& Rucinski(1995)]{kr95}
Kaluzny, J., \& Rucinski, S.~M.~1995, A\&AS, 114, 1

\bibitem[Koch \& McWilliam(2008)]{km08}
Koch, A., \& McWilliam, A.~2008, AJ, 135, 1551

\bibitem[Kraft \& Ivans(2003)]{ki03}
Kraft, R.~P., \& Ivans, I.~I.~2003, PASP, 115, 143

\bibitem[Kraft et al.(1992)]{ksl92}
Kraft, R.~P., Sneden, C., Langer, G.~E., \& Prosser, C.~F.~1992, AJ, 104, 645

\bibitem[Ku{\v c}inskas et al.(2009)]{klc09}
Ku{\v c}inskas, A., Ludwig, H.-G., Caffau, E., Steffen, M.~2009, Mem.~Soc.
Astron.~Ital., 80, 723

\bibitem[Landolt(1983)]{lan83}
Landolt, A.~U.~1983, AJ, 88, 439

\bibitem[Landolt(1992)]{lan92}
Landolt, A.~U.~1992, AJ, 104, 340

\bibitem[Lebreton, Fernandez, \& Lejeune(2001)]{lfj01}
Lebreton, Y., Fernandez, J., \& Lejeune, T.~2001, A\&A, 374, 540

\bibitem[Magic et al.(2010)]{msw10}
Magic, Z., Serenelli, A., Weiss, A., \& Chaboyer, B.~2010, ApJ, submitted

\bibitem[Marta et al.(2008)]{mfg08}
Marta, M., Formicola, A., Gy\"urky, Gy.~et al.~2008, Phys.~Rev.~C, 78, 022802

\bibitem[McCall(2004)]{mcc04}
McCall, M.~L.~2004, AJ, 128, 2144

\bibitem[Mel\'endez et al.(2010)]{mel10}
Mel\'endez, J., Schuster, W.~J., Silva, J.~S., Ram\'irez, I., Casagrande, L.,
\& Coelho, P.~2010, A\&A, submitted

\bibitem[Mel\'endez et al.(2006)]{msv06}
Mel\'endez, J., Shchukina, N.~G., Vasiljeva, I.~E., \& Ram\'irez, I.~2006,
ApJ, 642, 1082

\bibitem[Michaud et al.(2004)]{mrr04}
Michaud, G., Richard, O., Richer, J., \& VandenBerg, D.~A.~2004, ApJ, 606, 452

\bibitem[Milone et al.(2008)]{mbg08}
Milone, A.~P., Bedin, L.~R., Piotto, G., et al.~2008, ApJ, 673, 241

\bibitem[Milone et al.(2009)]{msp09}
Miline, A.~P., Stetson, P.~B., Piotto, G., Bedin, L.~R., Anderson, J., Cassisi,
S., \& Salaris, M.~2009, A\&A, 503, 755

\bibitem[Montgomery, Marschall, \& Janes(1993)]{mmj93}
Montgomery, K.~A., Marschall, L.~A., \& Janes, K.~A.~1993, AJ, 106, 181

\bibitem[Nissen et al.(2007)]{naa07}
Nissen, P.~E., Akerman, C., Asplund, M., Fabbian, D., Kerber, F., K\"auff,
H.~U., \& Pettini, M.~2007, A\&A, 469, 319

\bibitem[Nissen, Twarog, \& Crawford(1987)]{ntc87}
Nissen, P.~E., Twarog, B.~A., \& Crawford, D.~L.~1987, AJ, 93, 634

\bibitem[{\"O}nehag et al.(2009)]{oge09}
{\"O}nehag, A., Gustafsson, B., Eriksson, K., \& Edvardsson, B.~2009, A\&A,
 498, 527

\bibitem[Paulson, Sneden, \& Cochran(2003)]{psc03}
Paulson, D.~B., Sneden, C., \& Cochran, W.~D.~2003, AJ, 125, 3185

\bibitem[Pietrinferni et al.(2004)]{pcs04}
Pietrinferni, A., Cassisi, S., Salaris, M., \& Castelli, F.~2004, ApJ, 612, 168

\bibitem[Proffitt \& Michaud(1991)]{pm91}
Proffitt, C.~R., \& Michaud, G.~1991, ApJ, 371, 584

\bibitem[Ram\'irez \& Mel\'endez(2005)]{rm05}
Ram\'irez, I., \& Mel\'endez, J.~2005, ApJ, 626, 465

\bibitem[Randich et al.(2006)]{rsp06}
Randich, S., Sestito, P., Primas, F., Pallavicini, R., \& Pasquini, L.~2006,
A\&A, 450, 557

\bibitem[Reid(1993)]{rei93}
Reid, N.~1993, MNRAS, 265, 785

\bibitem[Richer, Fahlman, \& VandenBerg(1988)]{rfv88}
Richer, H.~B., Fahlman, G.~G., \& VandenBerg, D.~A.~1988, ApJ, 329, 187

\bibitem[Rood et al.(1999)]{rcp99}
Rood, R.~T., Carretta, E., Paltrinieri, B., et al.~1999, ApJ, 523, 752

\bibitem[Sandage, Lubin, \& VandenBerg(2003)]{slv03}
Sandage, A., Lubin, L.~M., \& VandenBerg, D.~A.~2003, PASP, 115, 1187

\bibitem[Sandquist(2004)]{san04}
Sandquist, E.~2004, MNRAS, 347, 104

\bibitem[Sandquist et al.(1996)]{sbs96}
Sandquist, E.~L., Bolte, M., Stetson, P.~B., \& Hesser, J.~E.~1996, ApJ, 470,
910

\bibitem[Sarajedini et al.(1999)]{svk99}
Sarajedini, A., von Hippel, T., Kozhurina-Platais, V., \& Demarque, P.~1999,
AJ, 118, 2894

\bibitem[Schlegel, Finkbeiner, \& Davis(1998)]{sfd98}
Schlegel, D., Finkbeiner, D.~P., \& Davis, M.~1998, ApJ, 500, 525

\bibitem[Skrutskie et al.(2006)]{scs06}
Skrutskie, M.~F., Cutri, R.~M., Stiening, R.,~et al.~2006, AJ, 131, 1163

\bibitem[Stetson(2000)]{ste00}
Stetson, P.~B.~2000, PASP, 112, 925

\bibitem[Stetson(2005)]{ste05}
Stetson, P.~B.~2005, PASP, 117, 563

\bibitem[Stetson(2009)]{ste09}
Stetson, P.~B.~2009, in The Ages of Stars, IAU Symp.~258, eds.~E.~E.~Mamajek,
D.~R.~Soderblom, \& R.~F.~G.~Wyse (Cambridge U.~Press, Cambridge), p.~197

\bibitem[Stetson, Bruntt, \& Grundahl(2003)]{sbg03}
Stetson, P.~B., Bruntt, H., \& Grundahl, F.~2003, PASP, 115, 413

\bibitem[Taylor \& Joner(1985)]{tj85}
Taylor, B.~J., \& Joner, M.~D.~1985, AJ, 96, 909

\bibitem[Tautvai{\u s}iene et al.(2000)]{tet00}
Tautvai{\u s}iene, G., Edvardsson, E., Tuominen, I., \& Ilyin, L.~2000, A\&A,
360, 495

\bibitem[VandenBerg(2000)]{van00}
VandenBerg, D.~A.~2000, ApJS, 129, 315

\bibitem[VandenBerg(2008)]{van08}
VandenBerg, D.~A.~2008, Physica Scripta, T133, 014026

\bibitem[VandenBerg, Bergbusch, \& Dowler(2006)]{vbd06}
VandenBerg, D.~A., Bergbusch, P.~A., \& Dowler, P.~D.~2006, ApJS, 162, 375

\bibitem[VandenBerg \& Clem(2003; hereafter VC03)]{vc03}
VandenBerg, D.~A., \& Clem, J.~L.~2003, AJ, 126, 778

\bibitem[VandenBerg et al.(2008)]{vee08}
VandenBerg, D.~A., Edvardsson, B., Eriksson, K., \& Gustafsson, B.~2008,
 ApJ, 675, 746 

\bibitem[VandenBerg et al.(2007)]{vge07}
VandenBerg, D.~A., Gustafsson, B., Edvardsson, E., Eriksson, K., \& Ferguson,
J.~2007, ApJ, 666, L105

\bibitem[VandenBerg \& Poll(1989)]{vp89}
VandenBerg, D.~A., \& Poll, H.~E.~1989, AJ, 98, 1451

\bibitem[VandenBerg et al.(2002)]{vrm02}
VandenBerg, D.~A., Richard, O., Michaud, G., \& Richer, J.~2002, ApJ, 571, 487

\bibitem[VandenBerg \& Stetson(2004)]{vs04}
VandenBerg, D.~A., \& Stetson, P.~B.~2004, PASP, 116, 997

\bibitem[van Leeuwen(2007)]{vl07}
van Leeuwen, F.~2007, A\&A, 474, 653

\bibitem[van Leeuwen(2009)]{vl09}
van Leeuwen, F.~2009, A\&A, 497, 209


\bibitem[Weiss(2008)]{wei08}
Weiss, A.~2008, Physica Scripta, T133, 014025

\bibitem[Weiss et al.(2007)]{wei07}
Weiss, A., Cassisi, S., Dotter, A., Han, Z., \& Lebreton, Y.~2007, in Stellar
Populations as Building Blocks of Galaxies, IAU Symp.~241, eds.~A.~Vazdekis \& 
R.~Peletier (Cambridge U.~Press: Cambridge), p.~28

\bibitem[Yong et al.(2009)]{ygd09}
Yong, D., Grundahl, F., D'Antona, F., Karakas, A.~I., Lattanzio, J.~C., \&
Norris, J.~E.~2009, ApJ, 695, 62

\bibitem[Zinn \& West(1984)]{zw84}
Zinn, R., \& West, M.~J.~1984, ApJS, 55, 45
 
\end{thebibliography}
\end{document}